\documentclass[aps,prd,amsmath,amssymb,showpacs,10pt]{revtex4-2}
\usepackage{graphicx}
\usepackage{color}
\usepackage{slashed}
\usepackage{enumerate}
\usepackage{physics}
\usepackage[font=footnotesize]{caption}
\usepackage[colorlinks,linkcolor=red,anchorcolor=magenta,citecolor=blue]{hyperref}
\allowdisplaybreaks
\begin{document}
\title{Combined study of the isospin-violating decay $D^{*}_{s}\to D_{s} \pi^0$ and radiative decay $D^*_s\to D_s\gamma$ with intermediate meson loops}
\author{Jun Wang$^{1,2}$}\email{junwang@ihep.ac.cn} 
\author{Qiang Zhao$^{1,2,3}$}\email{ zhaoq@ihep.ac.cn} 

\affiliation{
$^1$ Institute of High Energy Physics, Chinese Academy of Sciences, Beijing 100049, China \\
$^2$ University of Chinese Academy of Sciences, Beijing 100049, China \\
$^3$ Center for High Energy Physics, Henan Academy of Sciences, Zhengzhou 450046, China}

\begin{abstract}
    We carry out a combined study of the isospin-violating decay $D_{s}^{*} \to D_{s} \pi^{0}$ and radiative decay $D^*_s\to D_s\gamma$ in an effective Lagrangian approach by taking into account the corrections from the one-loop transitions. By distinguishing the transition mechanisms of the long-distance interactions through the intermediate meson loops from the short-distance interactions through the $\eta-\pi^{0}$ mixing at the tree level the isospin-violating decay $D_{s}^{*} \to D_{s} \pi^{0}$ can be well constrained. In our approach the higher order corrections to the isospin-violating effects can involve the intermediate $D^{(*)}$ and $K^{(*)}$ scatterings. We find that the contributions from the destructive interference of intermediate meson loops via $D^{(*)0}(c\bar{u}){K}^{(*)+}(u\bar{s})$ and $D^{(*)+}(c\bar{d}){K}^{(*)0}(d\bar{s})$ rescatterings are significant. Within the commonly accepted ultra-violet (UV) cutoff range we obtain the partial decay width  $\Gamma[D_{s}^{*} \to D_{s} \pi^{0}] = 9.92^{+0.76}_{-0.66}\,\mathrm{eV}$. This approach allows us to describe the $D_s^*$ radiative decay in the same framework via the vector meson dominance (VMD) model. We demonstrate that both the tree-level and one-loop transitions can be self-consistently determined if we adopt the experimental data for the branching ratio fraction of $D_{s}^{*} \to D_{s} \pi^{0}$ to $D^*_s\to D_s\gamma$. It then leads to a reliable estimate of the total decay width of $D_s^*$, i.e. $\Gamma_{\text{total}}(D_s^{*+})=170^{+ 13}_{-12}\,\mathrm{eV}$. 
\end{abstract}

\maketitle

\section{Introduction}
The study of isospin symmetry and its violations in hadronic decays provides critical insights into the dynamics of strong interactions in the non-perturbative regime. While isospin symmetry is approximately conserved in strong processes, its breaking, often arising from mass differences between up and down quarks and/or electromagnetic effects, offers a unique window into subtle aspects of hadron structure and decay mechanisms. Among such processes, the decay  $D_s^{*} \to D_s \pi^0$, which violates isospin symmetry, presents a compelling case for the detailed underlying dynamics. It is interesting to note that although the $D_s^*$ has been observed in experiment for a long time, its quantum number as the ground state of the charmed-strange vector meson was just measured very recently~\cite{BESIII:2023abz}. But its total width is still not  determined.

The strong and radiative decays of $D_s^*$ has attracted a lot of attention in the literature. Due to the limited phase space and isospin violation, the branching ratio of $D_s^*\to D_s\pi^0$ is much smaller than that of $D_s^*\to D_s\gamma$. While the radiative decay $D_s^*\to D_s\gamma$ via an $M1$ transition has a relatively well defined picture in the constituent quark model and there have been a lot of studies of the radiative transition~\cite{Aliev:1994nq,Cheng:1993kp,Choi:2007us,Deng:2013uca,Donald:2013sra,Fayyazuddin:1993eb,Goity:2000dk,Kamal:1992uv,Wang:2019mhm,Yu:2015xwa,Tran:2023hrn,Cheung:2015rya},
 the isospin-violating decay of $D_s^*\to D_s\pi^0$ has not yet been broadly investigated. Several recent works have been dedicated to this issue based on the $\eta-\pi^0$ mixing~\cite{Cho:1994zu,Ivanov:1998wn,Terasaki:2015eao,Cheung:2015rya}. In Ref.~\cite{Yang:2019cat} a heavy meson chiral perturbation theory calculation was presented with the $\mathcal{O}(p^3)$ loop corrections, and it was found that the $\mathcal{O}(p^3)$ corrections may actually be significant. 

On top of the $\eta-\pi^0$ mixing scenario, corrections from higher-order mechanisms are the focus of attention which is correlated with the total width question about $D_s^*$. In the framework of the heavy meson chiral perturbation theory~\cite{Yang:2019cat} it is shown that the next-to-leading-order loop diagrams of $\mathcal{O}(p^3)$ are associated with the counter term tree diagrams of the same order, where four undetermined low energy constants (LECs) are involved. Delicate considerations of estimating the $\mathcal{O}(p^3)$ tree diagram contributions are discussed in Ref.~\cite{Yang:2019cat} and the corrections can amount to about $30\%$ of the leading $\eta\pi^0$ mixing effects. While more constraints of the LECs should be sought, alternative approaches  should also be explored in order to elucidate the isospin breaking mechanism. 

In this work we propose an effective Lagrangian approach for calculating the higher-order corrections of the isospin-violating decay $D_s^*\to D_s\pi^0$. In addition to the leading tree-level amplitude from the $\eta-\pi^0$ mixing, we construct the higher-order corrections from the intermediate charmed meson $D^{(*)}$ and strange meson $K^{(*)}$ rescatterings. At the one-loop level the isospin-breaking can be categorized by two groups. One is the direct production of $\pi^0$ via the destructive interferences between the two loop amplitudes, namely between the $D^{(*)0}(c\bar{u}){K}^{(*)+}(u\bar{s})$ and $D^{(*)+}(c\bar{d}){K}^{(*)0}(d\bar{s})$. The other one is via the one-loop production of $\eta$ and then via the mixing of $\eta\pi^0$. In the latter case there are two corresponding loops involving the $u\bar{u}$ and $d\bar{d}$ productions, respectively, and they have a constructive phase first to produce $\eta$ and then couple to $\pi^0$ via the $\eta-\pi^0$ mixing. It should be noted that these two groups are not double-counting. 

It should also be noted that although the one-loop intermediate meson rescattering amplitudes do not simply count corrections of $\mathcal{O}(p^3)$, the leading one-loop amplitude provides corrections to the tree-level mixing term at $\mathcal{O}(m_\pi/m_{ex})$, where $m_{ex}$ is the exchanged meson mass between these two intermediate mesons, i.e. $D^{(*)}$ and $K^{(*)}$. We include all the ground-state vector and pseudoscalar meson rescatterings in the one-loop amplitude of which the vertex couplings can be well determined by either experimental measurements or heavy quark symmetry and chiral symmetry relations~\cite{Burdman:1992gh,Wise:1992hn,Casalbuoni:1996pg}. These amplitudes are unified by an overall cut-off parameter which is introduced with a form factor to regularize the loop integrals.  In this sense, the model uncertainties will be contained in the cutoff parameter, and the stability of the one-loop contributions can serve as a criterion for the model-dependent aspect of this approach. This method has been broadly applied to the studies of various isospin symmetry breaking processes~\cite{Li:2007au,Li:2008xm,Wang:2011yh,Guo:2010ak} and the helicity selection rule violation processes~\cite{Liu:2009vv,Liu:2010um,Zhang:2009kr} in the literature.

In the same framework we can then extend the study to the $D_s^*$ radiative decay via the vector meson dominance (VMD) model. The leading tree-level transition of $D_s^*\to D_s\gamma$ will go through the intermediate flavor neutral vector meson production followed by the vector meson conversion into the final-state photon. This can be categorized as the short-distance part of the radiative transition. Similar to the isospin-violating decay process, we can define the long-distance transition by the one-loop intermediate meson rescatterings where the vector meson will then convert into the final-state photon. Thus, these two dominant decay channels, i.e. $D_S^*\to D_s\gamma$ and $D_s^*\to D_s\pi^0$ can be combined together with the same set of parameters except for one different coupling parameter $g_{J/\psi D\bar{D}}$ introduced in $D_s^*\to D_s\gamma$. Their relative strength, i.e. the relative branching ratio fraction between  $D_S^*\to D_s\gamma$ and $D_s^*\to D_s\pi^0$, will provide a strong constraint on $g_{J/\psi D\bar{D}}$ in $D_s^*\to D_s\gamma$. Since this parameter can be determined by other independent processes, our calculation can examine the self-consistency of the formalism by comparing the constrained value for $g_{J/\psi D\bar{D}}$ with its commonly adopted values. Such a consistent constraint then allows us to extract the total width for $D_s^*$.

As follows, we first introduce our formalism in Sec. II. The numerical results and discussions will be presented in Sec. III. A brief summary and conclusion will be given in Sec. IV.

\section{Formalism}\label{sec2}
Our approach for the isospin-violating decay of $D^{*+}_{s}\to D_{s}^{+}\pi^0$ contains two main ingredients. The first one is the tree-level transition via the $\eta-\pi^0$ mixing. The second one is the loop transition via the intermediate $D^{(*)}$ and $K^{(*)}$ rescatterings. The electromagnetic transition contributions to the isospin-violating decay in $D^{*+}_{s}\to D_{s}^{+}\pi^0$ is believed to be negligibly small~\cite{Cho:1994zu,Yang:2019cat}. 

For $D_s^*\to D_s\gamma$, we calculate the leading $M_1$ transition in the VMD model where the $D_s^*$ first couples to $D_s\phi$ and $D_s J/\psi$, and then the vector mesons will convert into photon via the VMD mechanism. We also consider the one-loop correction in the same framework. The the intermediate $D^{(*)}$ and $K^{(*)}$ rescatterings will produce $D_s$ and the flavor neutral vector mesons and the latters will then convert into photon via the VMD mechanism. 

In this unified framework the effective Lagrangians are constructed by taking into account both the heavy quark symmetry and chiral symmetry~\cite{Burdman:1992gh,Wise:1992hn,Casalbuoni:1996pg}. The photon-vector-meson couplings can be determined by the leptonic decays of the vector mesons in $V\to e^+e^-$ with the help of the experimental measurements~\cite{ParticleDataGroup:2024cfk}.

\subsection{Effective Lagrangians}
To describe the strong couplings 
the following effective Lagrangians~\cite{Casalbuoni:1996pg,Cheng:2004ru,Wang:2012mf} are adopted :
\begin{equation}\label{Lagrangian-1}
  \begin{aligned}
    {\cal L} =& - ig_{\mathcal{D}^*\mathcal{D}\mathcal{P}}(\mathcal{D}^i\partial^\mu \mathcal{P}_{ij}
    \mathcal{D}_\mu^{*j\dagger}-\mathcal{D}_\mu^{*i}\partial^\mu \mathcal{P}_{ij}\mathcal{D}^{j\dagger})
    +\frac{1}{2}g_{\mathcal{D}^*\mathcal{D}^*\mathcal{P}}
    \varepsilon_{\mu\nu\alpha\beta}\,\mathcal{D}_i^{*\mu}\partial^\nu \mathcal{P}^{ij}
    {\overleftrightarrow \partial}{}^{\!\alpha} \mathcal{D}^{*\beta\dagger}_j \\
    -& ig_{\mathcal{DDV}} \mathcal{D}_i^\dagger {\overleftrightarrow \partial}_{\!\mu} \mathcal{D}^j(\mathcal{V}^\mu)^i_j
    -2f_{\mathcal{\mathcal{D^*DV}}} \epsilon_{\mu\nu\alpha\beta}
    (\partial^\mu V^\nu)^i_j
    (\mathcal{D}_i^\dagger{\overleftrightarrow \partial}{}^{\!\alpha} \mathcal{D}^{*\beta j}-\mathcal{D}_i^{*\beta\dagger}{\overleftrightarrow \partial}{}{\!^\alpha} \mathcal{D}^j)
   \\
    +& ig_{\mathcal{D^*D^*V}} \mathcal{D}^{*\nu\dagger}_i {\overleftrightarrow \partial}_{\!\mu} \mathcal{D}^{*j}_\nu(\mathcal{V}^\mu)^i_j
    +4if_{\mathcal{D^*D^*V}} \mathcal{D}^{*\dagger}_{i\mu}(\partial^\mu \mathcal{V}^\nu-\partial^\nu
   \mathcal{ V}^\mu)^i_j \mathcal{D}^{*j}_\nu, \\
  \end{aligned}
 \end{equation} 
where $\mathcal{D}$ and $\mathcal{D}^{*}$ represent the pseudoscalar and vector charm meson fields, respectively, i.e.
\begin{equation}
  \mathcal{D}=(D^{0},D^{+},D_{s}^{+}),\quad \mathcal{D}^{*}=(D^{*0},D^{*+},D_{s}^{*+}),
 \end{equation} 
$\mathcal{P}$ and $\mathcal{V}$ are $3\times 3$ matrices representing the pseudoscalar nonet and vector nonet meson fields \cite{Cao:2023gfv}
\begin{equation}
 \begin{aligned}
\mathcal{P}=
\begin{pmatrix}
    \frac{\sin \alpha_{P}\eta'+\cos \alpha_P \eta +\pi^{0}}{\sqrt{2}}& \pi^+ & K^+ \\
    \pi^- &\frac{\sin \alpha_{P}\eta'+\cos \alpha_P \eta -\pi^{0}}{\sqrt{2}}  & K^0 \\[1ex]
    K^-& {\bar K}^0 &\cos  \alpha_{P}\eta'-\sin  \alpha_P \eta \\
\end{pmatrix}\quad 
    \mathcal{V}=\begin{pmatrix}\frac{\rho^0+\omega} {\sqrt {2}}&\rho^+ & K^{*+} \\
    \rho^- & \frac {\omega-\rho^0} {\sqrt {2}} & K^{*0} \\[1ex]
    K^{*-}& {\bar K}^{*0} & \phi \\
  \end{pmatrix}.  \\
     \end{aligned}
     \end{equation} 
Specifically, for the process $D^{*+}_{s}\to D_{s}^{+}\eta$, the corresponding effective Lagrangian is
\begin{equation}
  \mathcal{L}_{D^{*}_{s}D_{s}\eta}=i g_{\mathcal{D}^{*}\mathcal{D}\mathcal{P}}\sin \alpha_P D_{s} (D^{*}_{s})_{\mu}\partial^{\mu}\eta\,.
 \end{equation} 
where $\alpha_P=40.6^\circ$ is the $\eta$-$\eta'$ mixing angle in the SU(3) flavor basis, and the value is an average from the Particle Data Group~\cite{ParticleDataGroup:2024cfk}.  For the $\eta \to \pi^{0}$ process, we use the $\eta-\pi^{0}$ mixing angle $\theta_{\eta \pi^{0}}$ which is given by the leading order chiral expansion~\cite{Gasser:1984gg}
     \begin{equation}
       \tan(2\theta_{\eta \pi^0})=\frac{\sqrt{3}}{2}\frac{m_{d}-m_{u}}{m_{s}-\hat{m}}.
      \end{equation} 
where $\hat{m}=(m_{u}+m_{d})/2$. The values of $m_{u}$, $m_{d}$, and $m_{s}$ are taken from the Particle Data Group \cite{ParticleDataGroup:2024cfk}. Since $\theta_{\eta \pi^0}$ is very small, we take
\begin{equation}
       \theta_{\eta \pi^0}\simeq \frac{\sqrt{3}}{4}\frac{m_{d}-m_{u}}{m_{s}-\hat{m}},
\end{equation}
as broadly adopted in the literature.

     For the light hadron vertices, we adopt the following effective Lagrangian:
   \begin{equation}
    \begin{aligned}
      \mathcal{L}_{VPP}=&i g_{VPP}\operatorname{Tr}[(\mathcal{P}\partial_{\mu}\mathcal{P}-\partial_{\mu}\mathcal{P} \mathcal{P})\mathcal{V}^{\mu}],\\ \mathcal{L}_{VVP}=&g_{VVP}\varepsilon_{\alpha \beta \mu \nu}\operatorname{Tr}[\partial^{\alpha}\mathcal{V}^{\mu}\partial^{\beta} \mathcal{V}^{\nu}\mathcal{P}],\\\mathcal{L}_{VVV}=&i g_{VVV}\operatorname{Tr}[(\partial_\mu V_{\nu}-\partial_{\nu}V_{\mu})V^{\mu}V^{\nu}]. \\
    \end{aligned}
   \end{equation} 
   For the coupling constants of $D$ mesons and light mesons, we adopt the following results \cite{Cheng:2004ru,Wang:2012mf,Wang:2011yh,Liu:2006dq,Zhang:2009kr}
  \begin{equation} 
    \begin{aligned}
& g_{D^{*}D^{*}\pi}=\frac{g_{D^{*}D\pi}}{\sqrt{m_{D}m_{D^{*}}}}=\frac{2g}{f_{\pi}},\quad g_{DDV}=g_{D^{*}D^{*}V}=\frac{\beta g_{V}}{\sqrt{2}},\quad f_{D^{*}DV}=\frac{f_{D^{*}D^{*}V}}{m_{D^{*}}}=\frac{\lambda g_{V}}{\sqrt{2}},\quad g_{D^{*}D_{s}K}=\sqrt{\frac{m_{D_s}}{m_{D}}}g_{D^{*}D \pi},\\[1ex]
    &g_{D_{s}^{*}DK}=\sqrt{\frac{m_{D_{s}^{*}}}{m_{D^{*}}}}g_{D^{*}D\pi},\quad g_{V}=\frac{m_{\rho}}{f_{\pi}},\quad g_{D^{*}D^{*}K}=\frac{g_{D^{*}DK}}{\sqrt{m_{D}m_{D^{*}}}}=\frac{2g}{f_{K}}\quad g_{D_{s}DV}=\sqrt{\frac{m_{Ds}}{m_{D}}}g_{DDV},\\[1ex]
    & g_{D^{*}D_{s}^{*}K^{*}}=\sqrt{\frac{m_{D_{s}^{*}}}{m_{D^{*}}}}g_{D^{*}D^{*}V},\quad f_{D^{*}D_{s}^{*}K^{*}}=\sqrt{\frac{m_{D^{*}_{s}}}{m_{D^{*}}}}f_{D^{*}D^{*}V},\quad \frac{g_{D^{*}_{s}D_{s}\mathcal{\eta}}}{\sqrt{m_{D^{*}_{s}}m_{D_{s}}}}=\frac{g_{D^{*}D\pi}\sin \alpha_P}{\sqrt{m_{D^{*} }m_{D}}}.
    \end{aligned}
    \end{equation} 
where $g=0.59,\beta=0.9,f_{\pi}=132\, \mathrm{MeV},f_{K}=155\,\mathrm{MeV}$and $\lambda=0.56\,\mathrm{GeV}^{-1}$ are adopted \cite{Cheng:2004ru,Isola:2003fh,Wang:2012mf,Zhang:2009kr,Wang:2011yh}.

The relative strengths and phases of the coupling constants for vector and scalar mesons can be determined by $SU(3)$ flavor symmetry relations, and expressed by overall coupling constants $g_{VVP}$ and $g_{VPP}$ \cite{Cao:2023gfv},
  \begin{equation}
    \begin{aligned}
      &g_{K^{*+}K^{*-}\pi^0}=-g_{K^{*0}\bar{K}^{*0}\pi^0}=\frac{1}{\sqrt{2}}g_{VVP},\\
      g_{K^{*0}\bar{K}^{0}\pi^0}=&-g_{K^{*0}\pi^0 \bar{K}^{0}}=g_{\bar{K}^{*0}\pi^0K^{0}}=-g_{\bar{K}^{*0}K^{0}\pi^0}=\frac{1}{\sqrt{2}}g_{VPP}, \\
      g_{K^{*+}\pi^0K^{-}}=&-g_{K^{*+}K^{-}\pi^0}=g_{K^{*-}K^{+}\pi^0}=-g_{K^{*-}\pi^0K^{+}}=\frac{1}{\sqrt{2}}g_{VPP}.
    \end{aligned}
   \end{equation} 	

  \subsection{Tree and loop transition amplitudes of\texorpdfstring{ $D^{*+}_{s}\to D_{s}^{+}\pi^{0}$}{DstoDspi} }
  \begin{figure}
%[!h]
      \centering
        \includegraphics[width=0.4\textwidth]{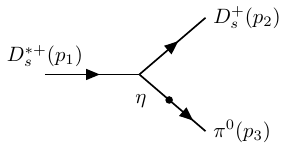}
        \caption{Schematic diagrams of the decay $D^{*+}_{s}\to D^{+}_{s}\pi^{0}$ via the $\eta-\pi^0$ mixing at tree level.}\label{fig:tree}
        \end{figure} 
Based on the effective Lagrangians in the previous section, we can derive the tree and loop amplitudes in order as follows. For the tree-level amplitude corresponding to Fig.\,\ref{fig:tree}, we have
          \begin{equation}
            i \mathcal{M}_{\text{tree}}=i  g_{D^{*}_{s}D_{s}\eta}\varepsilon_{D_{s}^{*}}\cdot p_{3}\theta_{\eta \pi^{0}}=i g_{\text{tree}}\varepsilon_{D^{*}_{s}}\cdot (p_2-p_{3}),
           \end{equation} 
where $g_{\text{tree}}\equiv g_{D^{*}_{s}D_{s}\eta}\theta_{\eta \pi^{0}}/2$ in our convention.

\begin{figure}
%[!h]
  \begin{minipage}[t]{0.32\textwidth}
    \begin{center}
      \includegraphics[width=1\textwidth]{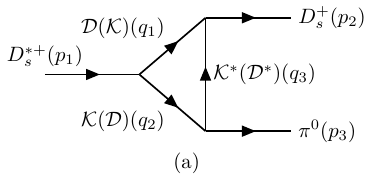}
    \end{center}
    \end{minipage}
     \begin{minipage}[t]{0.32\textwidth}
    \begin{center}
      \includegraphics[width=1\textwidth]{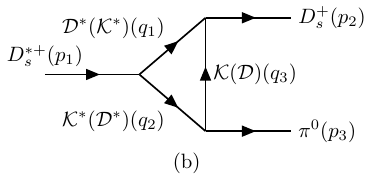}
    \end{center}
    \end{minipage}
     \begin{minipage}[t]{0.32\textwidth}
    \begin{center}
      \includegraphics[width=1\textwidth]{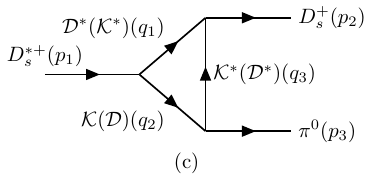}
    \end{center}
    \end{minipage}\\
     \begin{minipage}[t]{0.32\textwidth}
    \begin{center}
      \includegraphics[width=1\textwidth]{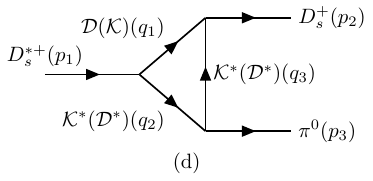}
    \end{center}
    \end{minipage}
     \begin{minipage}[t]{0.32\textwidth}
    \begin{center}
      \includegraphics[width=1\textwidth]{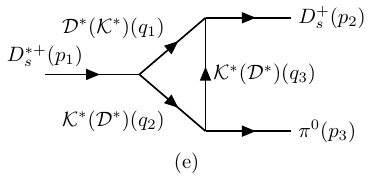}
    \end{center}
    \end{minipage}
    \caption{Schematic diagrams of the decay $D^{*+}_{s}\to D^{+}_{s}\pi^{0}$ via intermediate meson loops, where $\mathcal{D}=(D^{+},D^{0})$ and $\mathcal{K}=(K^{+},K^{0})$.}\label{fig:loop}
    \end{figure}   
For the loop-level amplitudes  without the $\eta-\pi^{0}$ mixing corresponding to Fig.\,\ref{fig:loop}, we denote the amplitude as $i\mathcal{M}(P_1,P_2,P_3)$, where $P_i$ represents the intermediate meson with momentum $q_i$, $\varepsilon^{s}$ is the polarization vector of the initial state $D^{*}_{s}$, and $\varepsilon^{i}$ is the polarization vector of the intermediate meson with momentum $q_{i}$. So the amplitudes corresponding to Fig.\,\ref{fig:loop} can be expressed as
    \begin{align}
      i \mathcal{M}_{a}(\mathcal{D},\mathcal{K},{\mathcal{K}}^{*})=&\int \frac{d^{4}q_{3}}{(2\pi)^{4}}\frac{g_{\mathcal{D}_{s}^{*}\mathcal{D}\mathcal{K}}q_{2}\cdot \varepsilon^{s}g_{\mathcal{D}_{s}\mathcal{D}{\mathcal{K}}^{*}}(q_1+p_2)_{\alpha}\qty(g^{\alpha \beta}-\frac{q_{3}^{\alpha}q_{3}^{\beta}}{m_{\bar{3}}^{2}})g_{{\mathcal{K}}^{*}\mathcal{K}\pi}(p_3+q_{2} )_{\beta}}{(q_1^{2}-m_1^{2})(q_2^{2}-m_2^{2})(q_3^{2}-m_3^{2})}\mathcal{F}(q_{i}^{2}), \\
      i \mathcal{M}_{a}(\mathcal{K},\mathcal{D},\mathcal{D}^{*})=&\int \frac{d^{4}q_{3}}{(2\pi)^{4}} \frac{g_{\mathcal{D}_{s}^{*}\mathcal{D}\mathcal{K}}g_{\mathcal{D}^{*}\mathcal{D}_{s}\mathcal{K}}g_{\mathcal{D}^{*}\mathcal{D}\pi} q_1\cdot  \varepsilon^{s} q_1^{\mu}\qty( g_{\mu\nu}-\frac{q_{3\mu}q_{3\nu}}{m_3^{2}})p_{3}^{\nu}}{(q_1^{2}-m_1^{2})(q_2^{2}-m_2^{2})(q_3^{2}-m_3^{2})}\mathcal{F}(q_{i}^{2}),\\
      i \mathcal{M}_{b}(\mathcal{D}^{*},\mathcal{K}^{*},\mathcal{K})    =&\int \frac{d^{4}q_{3}}{(2\pi)^{4}}\frac{(g_{\mathcal{D}_{s}^{*}\mathcal{D}^{*}\mathcal{K}^{*}}( p_1^{\mu}+q_1^{\mu})g^{\alpha \beta}-4f_{\mathcal{D}_{s}^{*}\mathcal{D}^{*}\mathcal{K}^{*}}(q_{2}^{\beta}g^{\alpha \mu}-q_{2}^{\alpha}g^{\beta \mu}))\varepsilon^{s}_{\alpha}g_{\mathcal{D}^{*}\mathcal{D}_{s}\mathcal{K}}q_3^{\nu}}{(q_1^{2}-m_1^{2})(q_2^{2}-m_2^{2})(q_3^{2}-m_3^{2})}\nonumber\\
      &\times g_{\mathcal{K}^{*}\mathcal{K}\pi}(p_3^{\lambda}-q_3^{\lambda})\qty(g_{\beta \nu}-\frac{q_{1\beta}q_{1\nu}}{m_1^{2}})\qty(g_{\mu\lambda}-\frac{q_{2\mu}q_{2\lambda}}{m_2^{2}})\mathcal{F}(q_{i}^{2}) ,\\
    i \mathcal{M}_{b}(\mathcal{K}^{*},\mathcal{D}^{*},\mathcal{D})
    =&-\int \frac{d^{4}q_{3}}{(2\pi)^{4}}\frac{(g_{\mathcal{D}_{s}^{*}\mathcal{D}^{*}\mathcal{K}^{*}}( p_1^{\mu}+q_2^{\mu})g^{\alpha \beta}-4f_{\mathcal{D}_{s}^{*}\mathcal{D}^{*}\mathcal{K}^{*}}(q_{1}^{\beta}g^{\alpha \mu}- q_{1}^{\alpha}g^{\beta \mu}))\varepsilon^{s}_{\alpha}g_{\mathcal{D}^{*}\mathcal{D} \pi}p_3^{\lambda}}{(q_1^{2}-m_1^{2})(q_2^{2}-m_2^{2})(q_3^{2}-m_3^{2})} \nonumber\\
    &
    \times g_{\mathcal{D}_{s}\mathcal{D}\mathcal{K}^{*}}(p_2+q_3)^{\nu}\qty(g_{\beta \nu}-\frac{q_{1\beta}q_{1\nu}}{m_1^{2}})\qty(g_{\mu\lambda}-\frac{q_{2\mu}q_{2\lambda}}{m_2^{2}})\mathcal{F}(q_{i}^{2}),
   \\
        i \mathcal{M}_{c}(\mathcal{D}^{*},\mathcal{K},\mathcal{K}^{*})
        =&-\int \frac{d^{4}q_{3}}{(2\pi)^{4}}\frac{4g_{\mathcal{D}^{*}_{s}\mathcal{D}^{*}\mathcal{K}}\varepsilon_{\mu\nu\alpha \beta}p_1^{\nu}\varepsilon^{s\mu} q_1^{\alpha} f_{\mathcal{D}^{*} \mathcal{D}_{s}\mathcal{K}^{*}}\varepsilon_{\kappa\sigma \zeta \rho}q_3^{\kappa}p_2^{\zeta}}{(q_1^{2}-m_1^{2})(q_2^{2}-m_2^{2})(q_3^{2}-m_3^{2})}\nonumber \\
        &
        \times g_{\mathcal{K}^{*}\mathcal{K}\pi}(p_{3\lambda}+ q_{2\lambda})\qty(g^{\beta \rho}-\frac{q_{1}^{\beta}q_{1}^{\rho}}{m_1^{2}})\qty(g^{\sigma\lambda}-\frac{q_{3}^{\sigma}q_{3}^{\lambda}}{m_3^{2}})\mathcal{F}(q_{i}^{2}), \\
          i \mathcal{M}_{c}(\mathcal{K}^{*},\mathcal{D},\mathcal{D}^{*})
        =&-\int \frac{d^{4}q_{3}}{(2\pi)^{4}}\frac{16 f_{\mathcal{D}^{*}_{s}\mathcal{D} \mathcal{K}^{*}}\varepsilon_{\alpha \beta \tau \mu} q_1^{\alpha}q_2^{\tau}\varepsilon^{s\mu}f_{\mathcal{D}^{*} \mathcal{D}_{s}\mathcal{K}^{*}}\varepsilon_{\kappa \rho \zeta \sigma}q_1^{\kappa}p_2^{\zeta}}{(q_1^{2}-m_1^{2})(q_2^{2}-m_2^{2})(q_3^{2}-m_3^{2})} \nonumber \\
        &
 \times g_{\mathcal{D}^{*}\mathcal{D}\pi} p_{3\lambda}\qty(g^{\beta \rho}-\frac{q_{1}^{\beta}q_{1}^{\rho}}{m_1^{2}})\qty(g^{\sigma\lambda}-\frac{q_{3}^{\sigma}q_{3}^{\lambda}}{m_3^{2}})\mathcal{F}(q_{i}^{2}),
 \\
        i \mathcal{M}_{d}(\mathcal{D},\mathcal{K}^{*},\mathcal{K}^{*})
        =&\int \frac{d^{4}q_{3}}{(2\pi)^{4}}\frac{4 f_{\mathcal{D}^{*}_{s}\mathcal{D} \mathcal{K}^{*}}\varepsilon_{\alpha \beta \tau \mu}q_2^{\alpha}q_1^{\tau}\varepsilon^{s\mu}g_{\mathcal{D} \mathcal{D}_{s}\mathcal{K}^{*}}(q_1+p_2)_{\sigma}}{(q_1^{2}-m_1^{2})(q_2^{2}-m_2^{2})(q_3^{2}-m_3^{2})}\nonumber  \\
        &
        \times g_{\mathcal{K}^{*}\mathcal{K}^{*}\pi}\varepsilon_{\kappa \zeta \rho \lambda}q_2^{\kappa}q_3^{\zeta}\qty(g^{\beta \rho}-\frac{q_{2}^{\beta}q_{2}^{\rho}}{m_2^{2}})\qty(g^{\sigma\lambda}-\frac{q_{3}^{\sigma}q_{3}^{\lambda}}{m_3^{2}})\mathcal{F}(q_{i}^{2}) ,\\
        i \mathcal{M}_{d}(\mathcal{K},\mathcal{D}^{*},\mathcal{D}^{*})
        =&-\int \frac{d^{4}q_{3}}{(2\pi)^{4}}\frac{g_{\mathcal{D}^{*}_{s}\mathcal{D}^{*}\mathcal{K}}\varepsilon_{\mu\nu\alpha \beta}p_1^{\nu}\varepsilon^{s\mu} q_2^{\alpha}g_{\mathcal{D}^{*}\mathcal{D}_{s}\mathcal{K}}q_1^{\sigma}}{(q_1^{2}-m_1^{2})(q_2^{2}-m_2^{2})(q_3^{2}-m_3^{2})}\nonumber \\
        &
        \times g_{\mathcal{D}^{*}\mathcal{D}^{*}\pi}\varepsilon_{\rho \zeta \kappa\lambda}p_3^{\zeta}q_3^{\kappa}\qty(g^{\beta \rho}-\frac{q_{2}^{\beta}q_{2}^{\rho}}{m_2^{2}})\qty(g^{\sigma\lambda}-\frac{q_{3}^{\sigma}q_{3}^{\lambda}}{m_3^{2}})\mathcal{F}(q_{i}^{2}),
     \\
          i \mathcal{M}_{e}(\mathcal{D}^{*},\mathcal{K}^{*},\mathcal{K}^{*})
        =&\int \frac{d^{4}q_{3}}{(2\pi)^{4}}\frac{4(g_{\mathcal{D}_{s}^{*}\mathcal{D}^{*}\mathcal{K}^{*}} (p_{1}+q_1)_{\mu}g_{\alpha \beta}-4f_{\mathcal{D}_{s}^{*}\mathcal{D}^{*}\mathcal{K}^{*}}(q_{2\beta}g_{\alpha \mu}- q_{2\alpha}g_{\beta \mu}))\varepsilon^{s\alpha}}{(q_1^{2}-m_1^{2})(q_2^{2}-m_2^{2})(q_3^{2}-m_3^{2})} \nonumber\\
      &\times  f_{\mathcal{D}^{*} \mathcal{D}_{s}\mathcal{K}^{*}}\varepsilon_{\kappa\sigma \zeta \rho}q_3^{\kappa} p_2^{\zeta}g_{\mathcal{K}^{*}\mathcal{K}^{*}\pi}\varepsilon_{\tau\delta\nu \lambda}q_2^{\tau}q_3^{\delta}\nonumber\\  &
        \times \qty(g^{\beta \rho}-\frac{q_{1}^{\beta}q_{1}^{\rho}}{m_1^{2}})\qty(g^{\mu\nu}-\frac{q_{2}^{\mu}q_{2}^{\nu}}{m_2^{2}})\qty(g^{\sigma\lambda}-\frac{q_{3}^{\sigma}q_{3}^{\lambda}}{m_3^{2}})\mathcal{F}(q_{i}^{2}),\\
        i \mathcal{M}_{e}(\mathcal{K}^{*},\mathcal{D}^{*},\mathcal{D}^{*})
        =&\int \frac{d^{4}q_{3}}{(2\pi)^{4}}\frac{4(g_{\mathcal{D}_{s}^{*}\mathcal{D}^{*}\mathcal{K}^{*}} (p_{1}+q_2)_{\beta}g_{\alpha \mu}-4f_{\mathcal{D}_{s}^{*}\mathcal{D}^{*}\mathcal{K}^{*}}(q_{1\mu}g_{\alpha \beta}- q_{1\alpha}g_{\beta \mu}))\varepsilon^{s\alpha}}{(q_1^{2}-m_1^{2})(q_2^{2}-m_2^{2})(q_3^{2}-m_3^{2})} \nonumber\\
        &\times f_{\mathcal{D}^{*} \mathcal{D}_{s}\mathcal{K}^{*}}\varepsilon_{\kappa \rho \zeta \sigma}q_1^{\kappa}p_2^{\zeta}g_{\mathcal{D}^{*}\mathcal{D}^{*}\pi}\varepsilon_{\nu\tau\delta\lambda}p_3^{\tau}q_3^{\delta}\nonumber\\
        &
        \times \qty(g^{\beta \rho}-\frac{q_{1}^{\beta}q_{1}^{\rho}}{m_1^{2}})\qty(g^{\mu\nu}-\frac{q_{2}^{\mu}q_{2}^{\nu}}{m_2^{2}})\qty(g^{\sigma\lambda}-\frac{q_{3}^{\sigma}q_{3}^{\lambda}}{m_3^{2}})\mathcal{F}(q_{i}^{2}).
        \end{align} 
In the above equations $\mathcal{F}(q_{i}^{2})$ is 
a form factor adopted for cutting off the ultraviolet divergence in the loop integrals, and has the following form
 \begin{equation}\label{formfactor}
  \mathcal{F}(p_{i}^{2})=\prod_{i}\qty(\frac{\Lambda_{i}^{2}-m_i^{2}}{\Lambda_{i}^{2}-p_{i}^{2}}),
  \end{equation} 
where $\Lambda_{i}\equiv m_{i}+\alpha \Lambda_{\text{QCD}}$ with the $m_i$ the mass of the $i$th internal particle, and the QCD energy scale $\Lambda_{\text{QCD}}=220\, \mathrm{MeV}$ with $\alpha=1\sim 2$ as the cutoff parameter~\cite{Guo:2010ak,Cao:2023csx}.

        \begin{figure}[!t]
          \begin{minipage}[t]{0.32\textwidth}
            \begin{center}
              \includegraphics[width=1\textwidth]{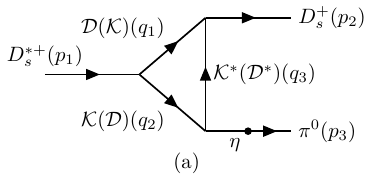}
            \end{center}
            \end{minipage}
             \begin{minipage}[t]{0.32\textwidth}
            \begin{center}
              \includegraphics[width=1\textwidth]{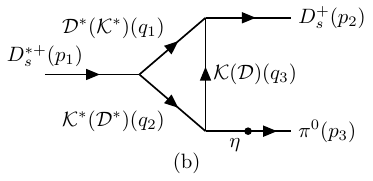}
            \end{center}
            \end{minipage}
             \begin{minipage}[t]{0.32\textwidth}
            \begin{center}
              \includegraphics[width=1\textwidth]{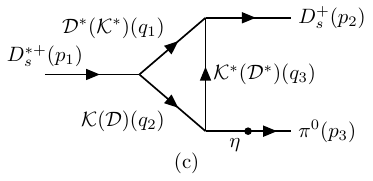}
            \end{center}
            \end{minipage}\\
             \begin{minipage}[t]{0.32\textwidth}
            \begin{center}
              \includegraphics[width=1\textwidth]{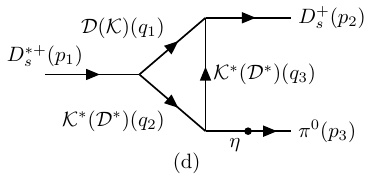}
            \end{center}
            \end{minipage}
             \begin{minipage}[t]{0.32\textwidth}
            \begin{center}
              \includegraphics[width=1\textwidth]{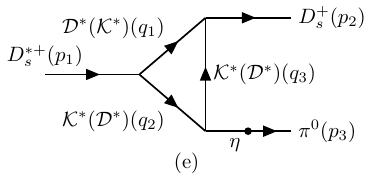}
            \end{center}
            \end{minipage}
            \caption{Schematic diagrams of the decay $D^{*+}_{s}\to D^{+}_{s}\pi^{0}$ via intermediate meson loops and $\eta-\pi$ mixing, where $\mathcal{D}=(D^{+},D^{0})$ and $\mathcal{K}=(K^{+},K^{0})$.}\label{fig:fig3}
            \end{figure}   
For the loop diagrams involving the $\eta-\pi^0$ mixing corresponding to Fig.\,\ref{fig:fig3}, the difference from the loop diagrams of  Fig.\,\ref{fig:loop} lies in the coupling constants related to $\eta$ and $\pi$. Namely, for those two isospin-related channels, their amplitudes will have a constructive phase in the $D_s^*\to D_s\eta$ transition, but will have a destructive phase in $D_s^*\to D_s\pi^0$.

The $\eta$ production amplitudes via the triangle loops will contribute to the $D_s^*\to D_s\pi^0$ channel via the $\eta-\pi^0$ mixing. We denote the amplitude as $i \mathcal{M}(P_1,P_2,P_3,\eta)$ and for each process in Fig.\,\ref{fig:fig3} their expressions are given below:
\begin{align}
    i \mathcal{M}_{a}(\mathcal{D},\mathcal{K},{\mathcal{K}}^{*},\eta)=&  i \mathcal{M}_{a}(\mathcal{D},\mathcal{K},{\mathcal{K}}^{*})\cdot \frac{g_{\mathcal{K}\mathcal{K}^{*}\eta}\theta_{\eta \pi^{0}}}{g_{\mathcal{K}\mathcal{K}^{*}\pi}},\quad 
    i \mathcal{M}_{a}(\mathcal{K},\mathcal{D},\mathcal{D}^{*},\eta)= i \mathcal{M}_{a}(\mathcal{K},\mathcal{D},\mathcal{D}^{*})\cdot \frac{g_{\mathcal{D}^{*}\mathcal{D} \eta}\theta_{\eta \pi^{0}}}{g_{\mathcal{D}^{*}\mathcal{D}\pi}},\\
    i \mathcal{M}_{b}(\mathcal{D}^{*},\mathcal{K}^{*},\mathcal{K},\eta)    =&i \mathcal{M}_{b}(\mathcal{D}^{*},\mathcal{K}^{*},\mathcal{K})\cdot \frac{g_{\mathcal{K}\mathcal{K}^{*}\eta}\theta_{\eta \pi^{0}}}{g_{\mathcal{K}\mathcal{K}^{*}\pi}},\quad
  i \mathcal{M}_{b}(\mathcal{K}^{*},\mathcal{D}^{*},\mathcal{D},\eta)
  =i \mathcal{M}_{b}(\mathcal{K}^{*},\mathcal{D}^{*},\mathcal{D})\cdot \frac{g_{\mathcal{D}^{*}\mathcal{D} \eta}\theta_{\eta \pi^{0}}}{g_{\mathcal{D}^{*}\mathcal{D}\pi}},\\
      i \mathcal{M}_{c}(\mathcal{D}^{*},\mathcal{K},\mathcal{K}^{*},\eta)
      =& i \mathcal{M}_{c}(\mathcal{D}^{*},\mathcal{K},\mathcal{K}^{*})\cdot \frac{g_{\mathcal{K}\mathcal{K}^{*}\eta}\theta_{\eta \pi^{0}}}{g_{\mathcal{K}\mathcal{K}^{*}\pi}},
\quad 
i \mathcal{M}_{c}(\mathcal{K}^{*},\mathcal{D},\mathcal{D}^{*},\eta)=i \mathcal{M}_{c}(\mathcal{K}^{*},\mathcal{D},\mathcal{D}^{*})\cdot \frac{g_{\mathcal{D}^{*}\mathcal{D} \eta}\theta_{\eta \pi^{0}}}{g_{\mathcal{D}^{*}\mathcal{D}\pi}},\\
      i \mathcal{M}_{d}(\mathcal{D},\mathcal{K}^{*},\mathcal{K}^{*},\eta)
      =& i \mathcal{M}_{d}(\mathcal{D},\mathcal{K}^{*},\mathcal{K}^{*})\cdot \frac{g_{\mathcal{K}^{*}\mathcal{K}^{*}\eta}\theta_{\eta \pi^{0}}}{g_{\mathcal{K}^{*}\mathcal{K}^{*}\pi}},\quad
      i \mathcal{M}_{d}(\mathcal{K},\mathcal{D}^{*},\mathcal{D}^{*},\eta)
      =i \mathcal{M}_{d}(\mathcal{K},\mathcal{D}^{*},\mathcal{D}^{*})\cdot \frac{g_{\mathcal{D}^{*}\mathcal{D}^{*} \eta}\theta_{\eta \pi^{0}}}{g_{\mathcal{D}^{*}\mathcal{D}^{*}\pi}},\\
        i \mathcal{M}_{e}(\mathcal{D}^{*},\mathcal{K}^{*},\mathcal{K}^{*},\eta)
      =& i \mathcal{M}_{e}(\mathcal{D}^{*},\mathcal{K}^{*},\mathcal{K}^{*})\cdot \frac{g_{\mathcal{K}^{*}\mathcal{K}^{*}\eta}\theta_{\eta \pi^{0}}}{g_{\mathcal{K}^{*}\mathcal{K}^{*}\pi}},\quad 
      i \mathcal{M}_{e}(\mathcal{K}^{*},\mathcal{D}^{*},\mathcal{D}^{*},\eta)
      =i \mathcal{M}_{e}(\mathcal{K}^{*},\mathcal{D}^{*},\mathcal{D}^{*})\cdot \frac{g_{\mathcal{D}^{*}\mathcal{D}^{*} \eta}\theta_{\eta \pi^{0}}}{g_{\mathcal{D}^{*}\mathcal{D}^{*}\pi}} \ .
      \end{align} 

As mentioned earlier, there does not exist a simple power counting in the framework of the effective Lagrangian approach. In order to show the relation between the tree and loop amplitudes, we take a typical loop amplitude, e.g.  Fig.\,\ref{fig:loop} (a), to demonstrate that it counts  $\mathcal{O}(m_\pi/m_3)$ with $m_3$ the mass of the exchanged meson in the triangle loop. Since the masses of the intermediate $D$ and final $D_s$ are comparable, and the threshold of $DK$ is about 253 MeV which is smaller than the masses of the exchanged mesons. By approximation we assume that the main contribution of the loop amplitude is from the kinematic region that the intermediate $DK$ are nearly on shell and with a non-relativistic velocity $v$, while the exchanged mesons are highly off-shell with a small value of $|t|$.  With the treatment $1/(t-m_3^2)\simeq -1/m_3^2$ for the exchanged meson the loop amplitude scales as $(v^5/v^4)\times p_\pi\cdot p_{D_s}/m_3^2\simeq v E_{D_s} E_\pi/m_3^2$. Taking into account the isospin breaking, the cancellation between these two isospin amplitudes scales as $v  E_{D_s} E_\pi (1/m_{K^{*\pm}}^2-1/m_{K^{*0}}^2)\simeq v E_\pi\delta_{K^*}/m_{K^{*}}^2\simeq v (m_\pi/m_{K^*}) (\delta_{K^*}/m_{K^*})$, with $\delta_{K^*}\equiv m_{K^{*0}}-m_{K^{*\pm}}$. Although we do not include the coupling constants, the factor $\delta_{K^*}/m_{K^*}$ represents the source of the isospin violation from the mass difference between the $u$ and $d$ quark. In contrast, the factor $m_\pi/m_{K^*}$ indicates the suppression of the loop amplitudes of  Fig.\,\ref{fig:loop} (a) relative to the tree-level amplitude as a higher-order correction. Although the above analysis is qualitative, the numerical results with a reasonable cut-off for the ultra-violet (UV) divergence can provide a reliable estimate of the isospin-breaking corrections.

From the above amplitudes, we can obtain the total loop amplitude by summing up the contributions from the different loop diagrams. The decay $D^{*+}_{s}\to D^{+}_{s}\pi^0$ is a $VPP$ type decay process, and we can always parametrize the total loop amplitude as
 \begin{equation}
   i \mathcal{M}_{\text{loop}}= i g_{\text{loop}}\varepsilon_{D^{*+}_{s}}\cdot(p_2-p_3).
  \end{equation} 
Taking into account the tree amplitude, the total decay amplitude can be expressed as
  \begin{equation}
   i \mathcal{M}_{D^{*+}_{s}\to D^{+}_{s}\pi^0}=i(g_{\text{tree}}+g_{\text{loop}})\varepsilon_{D^{*+}_{s}}\cdot (p_{D^{+}_{s}}-p_{\pi^{0}})\equiv ig_{\text{total}}\varepsilon_{D^{*+}_{s}}\cdot (p_{D^{+}_{s}}-p_{\pi^0}) \ ,
   \end{equation} 
where $g_{\text{total}}$ is the effective coupling for $D^{*+}_{s}\to D^{+}_{s}\pi^0$. Then, the corresponding partial decay width is
\begin{equation}
  \Gamma_{D^{*+}_{s}\to D_{s}^{+}\pi^0}= \frac{\abs{\mathbf{p}}^{3}g_{\text{total}}^2}{6\pi M^{2}_{V}}.
 \end{equation} 
\subsection{Tree and loop amplitudes of \texorpdfstring{$D^{*+}_{s}\to D_{s}^{+}\gamma$}{Ds to Ds gamma} in the VMD model}
\begin{figure}[!ht]
    \centering
    \includegraphics[width=0.4\textwidth]{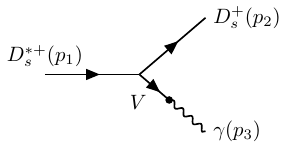}
    \caption{Tree-level diagram of $D^{*}_{s}\to D_{s}\gamma$ in the VMD model.}\label{fig:vmd_tree}
\end{figure}

Similar to the isospin-violating decay $D^{*+}_{s}\to D_{s}^{+}\pi^{0}$, we can also calculate the radiative decay $D^{*+}_{s}\to D^{+}_{s}\gamma$. The radiative decay can be described using the VMD model. The tree-level diagram is shown in Fig.~\ref{fig:vmd_tree}.
    The tree-level amplitude for the process $D^{*}_{s}\to D_{s}\gamma$ can be expressed as
\begin{equation}
  i \mathcal{M}_{\text{tree}}^{\gamma}=i g_{\text{tree}}^{\gamma}({\gamma})\varepsilon_{\mu\nu\alpha \beta}p_{1}^{\mu}p_3^{\nu}\varepsilon^{\alpha}_{1}\varepsilon^{\beta}_{3} \ ,
 \end{equation} 
where $g_{\text{tree}}^{\gamma}$ can be calculated using the VMD model,
 \begin{equation}
  g_{\text{tree}}^{\gamma}=i g_{D^{*}_{s}D_{s}V}\frac{em^{2}_{V}}{f_{V}}G_{V}
  \end{equation} 
with
  \begin{equation}
    G_{V}\equiv \frac{-i}{p^{2}_{\gamma}-m^{2}_{V}+i m_{V}\Gamma_{V}}=\frac{-i}{-m^{2}_{V}+i m_{V}\Gamma_{V}}.
   \end{equation} 
In the above equations $V= \rho,\omega,\phi $, and $e/f_{V}$ can be determined by experimental data of $\Gamma_{V\to e^{+}e^{-}}$:
   \begin{equation}
     \frac{e}{f_{V}}=\qty[\frac{3 \Gamma_{V\to e^{+}e^{-}}}{2\alpha_e\abs{\mathbf{p}_{e}}}]^{\frac{1}{2}}
    \end{equation} 
    where $\abs{\mathbf{p}_{e}}$ is the momentum of the final-state particle in the center-of-mass frame, and $\alpha_e=\frac{1}{137}$ is the fine-structure constant. 
To match the coupling constants in HQEFT, we have
   \begin{equation}     g_{D^{*}_{s}D_{s}V}=4f_{D^{*}_{s}D_{s}V}.
    \end{equation} 
    For the tree level $V=\phi, \ J/\psi$, the coupling constant can be extracted:
\begin{equation}
    g_{D^{*}_{s}D_{s}\gamma}=i \qty(g_{D^{*}_{s}D_{s}\phi}\frac{em^{2}_{\phi}}{f_{\phi}}G_{\phi}+g_{D^{*}_{s}D_{s}\psi}\frac{em^{2}_{\psi}}{f_{\psi}}G_{\psi}) \ ,
 \end{equation} 
with $g_{D^{*}_{s}D_{s}\psi}=2g_{J\psi D \bar{D}}/ \tilde{M}, \tilde{M}=\sqrt{M_{D_{s}^{*}}M_{D_{s}}}$\cite{Zhang:2009kr}.

\begin{figure}[!t]
    \begin{minipage}[t]{0.32\textwidth}
      \begin{center}
        \includegraphics[width=1\textwidth]{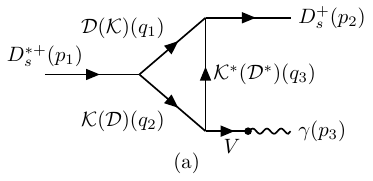}
            \end{center}
            \end{minipage}
             \begin{minipage}[t]{0.32\textwidth}
            \begin{center}
        \includegraphics[width=1\textwidth]{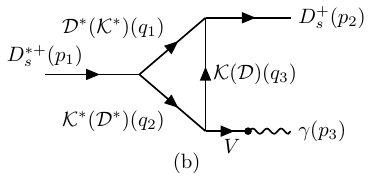}
            \end{center}
            \end{minipage}
             \begin{minipage}[t]{0.32\textwidth}
            \begin{center}
        \includegraphics[width=1\textwidth]{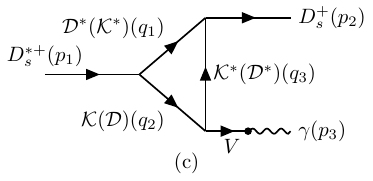}
            \end{center}
            \end{minipage}\\
             \begin{minipage}[t]{0.32\textwidth}
            \begin{center}
        \includegraphics[width=1\textwidth]{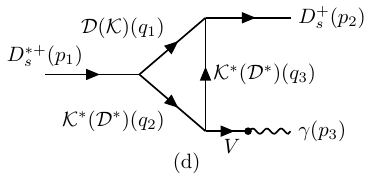}
            \end{center}
            \end{minipage}
             \begin{minipage}[t]{0.32\textwidth}
            \begin{center}
        \includegraphics[width=1\textwidth]{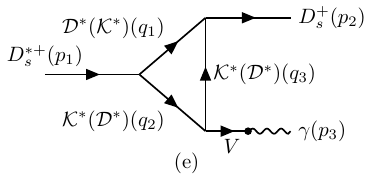}
            \end{center}
            \end{minipage}
            \begin{minipage}[t]{0.32\textwidth}
        \begin{center}
          \includegraphics[width=1\textwidth]{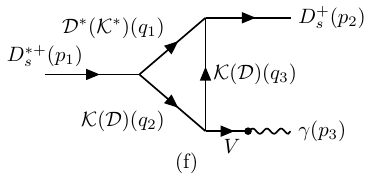}
        \end{center}
        \end{minipage}
            \caption{Schematic diagrams of the decay $D^{*+}_{s}\to D^{+}_{s}\gamma$ via intermediate meson loops in VMD model where $V=\rho ,\omega,\phi$.}\label{fig:vmdloop}
            \end{figure}   

The schematic diagrams for the loop contributions are shown in Fig.~\ref{fig:vmdloop}. Similar to the process $D^{*+}_{s}\to D^{+}_{s} \pi^{0}$, the corresponding amplitudes can be derived using the VMD model:
\begin{align}i \mathcal{M}_{a}(\mathcal{D},\mathcal{K},{\mathcal{K}}^{*},\gamma)=&\int \frac{d^{4}q_{3}}{(2\pi)^{4}}\frac{g_{\mathcal{K}\mathcal{K}^{*}\gamma}g_{\mathcal{D}_{s}^{*}\mathcal{D}\mathcal{K}}g_{\mathcal{D}_{s}\mathcal{D}{\mathcal{K}}^{*}}q_{2}\cdot \varepsilon^{s}(q_1+p_2)_{\alpha}\qty(g^{\alpha \beta}-\frac{q_{3}^{\alpha}q_{3}^{\beta}}{m_{\bar{3}}^{2}})\varepsilon_{\beta \sigma\mu\nu}\varepsilon_{3}^{\sigma}q_{3}^{\mu}p_{3}^{\nu}}{(q_1^{2}-m_1^{2})(q_2^{2}-m_2^{2})(q_3^{2}-m_3^{2})}\mathcal{F}(q_{i}^{2}), 
   \\i \mathcal{M}_{a}(\mathcal{K},\mathcal{D},\mathcal{D}^{*},\gamma)=&\int \frac{d^{4}q_{3}}{(2\pi)^{4}} \frac{g_{\mathcal{D}_{s}^{*}\mathcal{D}\mathcal{K}}g_{\mathcal{D}^{*}\mathcal{D}_{s}\mathcal{K}}g_{\mathcal{D}^{*}\mathcal{D}\gamma} q_1\cdot  \varepsilon^{s} q_1^{\mu}\qty( g_{\mu\nu}-\frac{q_{3\mu}q_{3\nu}}{m_3^{2}})\varepsilon_{\nu \sigma \alpha \beta}\varepsilon_{3}^{\sigma}q_{3}^{\alpha}p_{3}^{\beta}}{(q_1^{2}-m_1^{2})(q_2^{2}-m_2^{2})(q_3^{2}-m_3^{2})}\mathcal{F}(q_{i}^{2}),
\\i \mathcal{M}_{b}(\mathcal{D}^{*},\mathcal{K}^{*},\mathcal{K},\gamma)    =&\int \frac{d^{4}q_{3}}{(2\pi)^{4}}\frac{g_{\mathcal{D}^{*}\mathcal{D}_{s}\mathcal{K}}g_{\mathcal{K}^{*}\mathcal{K}\gamma}(g_{\mathcal{D}_{s}^{*}\mathcal{D}^{*}\mathcal{K}^{*}}( p_1^{\mu}+q_1^{\mu})g^{\alpha \beta}-4f_{\mathcal{D}_{s}^{*}\mathcal{D}^{*}\mathcal{K}^{*}}(q_{2}^{\beta}g^{\alpha \mu}-q_{2}^{\alpha}g^{\beta \mu}))\varepsilon^{s}_{\alpha}q_3^{\nu}}{(q_1^{2}-m_1^{2})(q_2^{2}-m_2^{2})(q_3^{2}-m_3^{2})}\nonumber\\
    &\times \varepsilon_{\lambda \sigma \kappa \rho}\varepsilon_{3}^{\sigma}q_{3}^{\kappa}p_{3}^{\rho}\qty(g_{\beta \nu}-\frac{q_{1\beta}q_{1\nu}}{m_1^{2}})\qty(g_{\mu\lambda}-\frac{q_{2\mu}q_{2\lambda}}{m_2^{2}})\mathcal{F}(q_{i}^{2}) ,
 \\i \mathcal{M}_{b}(\mathcal{K}^{*},\mathcal{D}^{*},\mathcal{D},\gamma)
  =&-\int \frac{d^{4}q_{3}}{(2\pi)^{4}}\frac{g_{\mathcal{D}_{s}\mathcal{D}\mathcal{K}^{*}}g_{\mathcal{D}^{*}\mathcal{D}\gamma}(g_{\mathcal{D}_{s}^{*}\mathcal{D}^{*}\mathcal{K}^{*}}( p_1^{\mu}+q_2^{\mu})g^{\alpha \beta}-4f_{\mathcal{D}_{s}^{*}\mathcal{D}^{*}\mathcal{K}^{*}}(q_{1}^{\beta}g^{\alpha \mu}- q_{1}^{\alpha}g^{\beta \mu}))\varepsilon^{s}_{\alpha}}{(q_1^{2}-m_1^{2})(q_2^{2}-m_2^{2})(q_3^{2}-m_3^{2})} \nonumber\\
  &
  \times \varepsilon_{\lambda \sigma \kappa \rho}\varepsilon_{3}^{\sigma}q_{3}^{\kappa}p_{3}^{\rho}(p_2+q_3)^{\nu}\qty(g_{\beta \nu}-\frac{q_{1\beta}q_{1\nu}}{m_1^{2}})\qty(g_{\mu\lambda}-\frac{q_{2\mu}q_{2\lambda}}{m_2^{2}})\mathcal{F}(q_{i}^{2}),
   \\i \mathcal{M}_{c}(\mathcal{D}^{*},\mathcal{K},\mathcal{K}^{*},\gamma)
      =&-\int \frac{d^{4}q_{3}}{(2\pi)^{4}}\frac{4g_{\mathcal{K}\mathcal{K}^{*}\gamma}f_{\mathcal{D}^{*} \mathcal{D}_{s}\mathcal{K}^{*}}g_{\mathcal{D}^{*}_{s}\mathcal{D}^{*}\mathcal{K}}\varepsilon_{\mu\nu\alpha \beta}p_1^{\nu}\varepsilon^{s\mu} q_1^{\alpha} \varepsilon_{\kappa\sigma \zeta \rho}q_3^{\kappa}p_2^{\zeta}}{(q_1^{2}-m_1^{2})(q_2^{2}-m_2^{2})(q_3^{2}-m_3^{2})}\nonumber \\
      &
      \times \varepsilon_{\lambda \tau \kappa \rho}\varepsilon_{3}^{\tau}q_{3}^{\kappa}p_{3}^{\rho}\qty(g^{\beta \rho}-\frac{q_{1}^{\beta}q_{1}^{\rho}}{m_1^{2}})\qty(g^{\sigma\lambda}-\frac{q_{3}^{\sigma}q_{3}^{\lambda}}{m_3^{2}})\mathcal{F}(q_{i}^{2}), 
       \\i \mathcal{M}_{c}(\mathcal{K}^{*},\mathcal{D},\mathcal{D}^{*},\gamma)
      =&-\int \frac{d^{4}q_{3}}{(2\pi)^{4}}\frac{16 g_{\mathcal{D}^{*}\mathcal{D}\gamma}f_{\mathcal{D}^{*} \mathcal{D}_{s}\mathcal{K}^{*}}f_{\mathcal{D}^{*}_{s}\mathcal{D} \mathcal{K}^{*}}\varepsilon_{\alpha \beta \tau \mu} q_1^{\alpha}q_2^{\tau}\varepsilon^{s\mu}\varepsilon_{\kappa \rho \zeta \sigma}q_1^{\kappa}p_2^{\zeta}}{(q_1^{2}-m_1^{2})(q_2^{2}-m_2^{2})(q_3^{2}-m_3^{2})} \nonumber \\
      &
\times \varepsilon_{\lambda \tau \nu \delta}\varepsilon_{3}^{\tau}q_{3}^{\nu}p_{3}^{\delta}\qty(g^{\beta \rho}-\frac{q_{1}^{\beta}q_{1}^{\rho}}{m_1^{2}})\qty(g^{\sigma\lambda}-\frac{q_{3}^{\sigma}q_{3}^{\lambda}}{m_3^{2}})\mathcal{F}(q_{i}^{2}),
      \\i \mathcal{M}_{d}(\mathcal{D},\mathcal{K}^{*},\mathcal{K}^{*},\gamma)
      =&\int \frac{d^{4}q_{3}}{(2\pi)^{4}}\frac{4g_{\mathcal{K}\mathcal{K}^{*}\gamma}g_{\mathcal{D} \mathcal{D}_{s}\mathcal{K}^{*}} f_{\mathcal{D}^{*}_{s}\mathcal{D} \mathcal{K}^{*}}\varepsilon_{\alpha \beta \tau \mu}q_2^{\alpha}q_1^{\tau}\varepsilon^{s\mu}(q_1+p_2)_{\sigma}}{(q_1^{2}-m_1^{2})(q_2^{2}-m_2^{2})(q_3^{2}-m_3^{2})}\nonumber  \\
      &
      \times (\varepsilon_{3\rho}p_{3\lambda}-\varepsilon_{3\lambda}p_{3\rho})\qty(g^{\beta \rho}-\frac{q_{2}^{\beta}q_{2}^{\rho}}{m_2^{2}})\qty(g^{\sigma\lambda}-\frac{q_{3}^{\sigma}q_{3}^{\lambda}}{m_3^{2}})\mathcal{F}(q_{i}^{2}) ,
      \\i \mathcal{M}_{d}(\mathcal{K},\mathcal{D}^{*},\mathcal{D}^{*},\gamma)
      =&-\int \frac{d^{4}q_{3}}{(2\pi)^{4}}\frac{g_{\mathcal{D}^{*}_{s}\mathcal{D}^{*}\mathcal{K}}g_{\mathcal{D}^{*}\mathcal{D}_{s}\mathcal{K}}(g_{\mathcal{D}^{*}\mathcal{D}^{*}\gamma}(q_{2}+p_3)\cdot \varepsilon_{3}g_{\rho\lambda}-f_{\mathcal{D}^{*}\mathcal{D}^{*}\gamma}(\varepsilon_{3\rho}p_{3\lambda}-\varepsilon_{3\lambda}p_{3\rho}))}{(q_1^{2}-m_1^{2})(q_2^{2}-m_2^{2})(q_3^{2}-m_3^{2})}\nonumber \\
      &
      \times\varepsilon_{\mu\nu\alpha \beta}p_1^{\nu}\varepsilon^{s\mu} q_2^{\alpha} q_1^{\sigma}\qty(g^{\beta \rho}-\frac{q_{2}^{\beta}q_{2}^{\rho}}{m_2^{2}})\qty(g^{\sigma\lambda}-\frac{q_{3}^{\sigma}q_{3}^{\lambda}}{m_3^{2}})\mathcal{F}(q_{i}^{2}),
      \\i \mathcal{M}_{e}(\mathcal{D}^{*},\mathcal{K}^{*},\mathcal{K}^{*},\gamma)
      =&\int \frac{d^{4}q_{3}}{(2\pi)^{4}}\frac{4(g_{\mathcal{D}_{s}^{*}\mathcal{D}^{*}\mathcal{K}^{*}} (p_{1}+q_1)_{\mu}g_{\alpha \beta}-4f_{\mathcal{D}_{s}^{*}\mathcal{D}^{*}\mathcal{K}^{*}}(q_{2\beta}g_{\alpha \mu}- q_{2\alpha}g_{\beta \mu}))\varepsilon^{s\alpha}}{(q_1^{2}-m_1^{2})(q_2^{2}-m_2^{2})(q_3^{2}-m_3^{2})} \nonumber\\
    &\times  f_{\mathcal{D}^{*} \mathcal{D}_{s}\mathcal{K}^{*}}\varepsilon_{\kappa\sigma \zeta \rho}q_3^{\kappa} p_2^{\zeta}g_{\mathcal{K}\mathcal{K}^{*}\gamma}(\varepsilon_{3\nu}p_{3\lambda}-\varepsilon_{3\lambda}p_{3\nu})\nonumber\\  &
      \times \qty(g^{\beta \rho}-\frac{q_{1}^{\beta}q_{1}^{\rho}}{m_1^{2}})\qty(g^{\mu\nu}-\frac{q_{2}^{\mu}q_{2}^{\nu}}{m_2^{2}})\qty(g^{\sigma\lambda}-\frac{q_{3}^{\sigma}q_{3}^{\lambda}}{m_3^{2}})\mathcal{F}(q_{i}^{2}),
    \\i \mathcal{M}_{e}(\mathcal{K}^{*},\mathcal{D}^{*},\mathcal{D}^{*},\gamma)
      =&\int \frac{d^{4}q_{3}}{(2\pi)^{4}}\frac{4 f_{\mathcal{D}^{*} \mathcal{D}_{s}\mathcal{K}^{*}}(g_{\mathcal{D}_{s}^{*}\mathcal{D}^{*}\mathcal{K}^{*}} (p_{1}+q_2)_{\beta}g_{\alpha \mu}-4f_{\mathcal{D}_{s}^{*}\mathcal{D}^{*}\mathcal{K}^{*}}(q_{1\mu}g_{\alpha \beta}- q_{1\alpha}g_{\beta \mu}))\varepsilon^{s\alpha}}{(q_1^{2}-m_1^{2})(q_2^{2}-m_2^{2})(q_3^{2}-m_3^{2})} \nonumber\\
      &\times\varepsilon_{\kappa \rho \zeta \sigma}q_1^{\kappa}p_2^{\zeta} (g_{\mathcal{D}^{*}\mathcal{D}^{*}\gamma}(q_{2}+p_3)\cdot \varepsilon_{3}g_{\nu\lambda}-f_{\mathcal{D}^{*}\mathcal{D}^{*}\gamma}(\varepsilon_{3\nu}p_{3\lambda}-\varepsilon_{3\lambda}p_{3\nu}))\nonumber\\
      &
      \times \qty(g^{\beta \rho}-\frac{q_{1}^{\beta}q_{1}^{\rho}}{m_1^{2}})\qty(g^{\mu\nu}-\frac{q_{2}^{\mu}q_{2}^{\nu}}{m_2^{2}})\qty(g^{\sigma\lambda}-\frac{q_{3}^{\sigma}q_{3}^{\lambda}}{m_3^{2}})\mathcal{F}(q_{i}^{2}).
    \\i \mathcal{M}_{f}(\mathcal{D}^{*},\mathcal{K},\mathcal{K},\gamma)
        =&-\int \frac{d^{4}q_{3}}{(2\pi)^{4}}\frac{g_{\mathcal{D}^{*}_{s}\mathcal{D}^{*}\mathcal{K}}g_{\mathcal{D}^{*}\mathcal{D}\mathcal{K}}g_{\mathcal{K}\mathcal{K}\gamma}\varepsilon_{\mu\nu\alpha \beta}p_1^{\nu}\varepsilon^{s\mu} q_1^{\alpha}p_{2}^{\rho} \varepsilon_{3}\cdot (q_2+q_3)\qty(g^{\beta \rho}-\frac{q_{1}^{\beta}q_{1}^{\rho}}{m_1^{2}})}{(q_1^{2}-m_1^{2})(q_2^{2}-m_2^{2})(q_3^{2}-m_3^{2})}\mathcal{F}(q_{i}^{2}),
      \\i \mathcal{M}_{f}(\mathcal{K}^{*},\mathcal{D},\mathcal{D},\gamma)
        =&-\int \frac{d^{4}q_{3}}{(2\pi)^{4}}\frac{4 f_{\mathcal{D}^{*}_{s}\mathcal{D} \mathcal{K}^{*}}g_{\mathcal{D}\mathcal{D}\mathcal{K}^{*}}g_{\mathcal{D}\mathcal{D}\gamma}\varepsilon_{\alpha \beta \tau \mu} q_1^{\alpha}q_2^{\tau}\varepsilon^{s\mu}(q_1+p_2)_{\rho}\varepsilon_{3}\cdot (q_2+q_3)\qty(g^{\beta \rho}-\frac{q_{1}^{\beta}q_{1}^{\rho}}{m_1^{2}})}{(q_1^{2}-m_1^{2})(q_2^{2}-m_2^{2})(q_3^{2}-m_3^{2})} \mathcal{F}(q_{i}^{2}).
        \end{align}
In the above equations the corresponding radiative decay coupling constants can be calculated using the VMD model \cite{Cao:2023gfv}, the same form factor as Eq.~(\ref{formfactor}) is adopted. The relevant expressions are as follows: 
 \begin{align}  g_{K^+K^{*+}\gamma}=&i(g_{\rho^0K^{*+}K^-}\frac{em_{\rho^0}^2}{f_{\rho^0}}G_{\rho^0}+g_{\omega K^{*+}K^-}\frac{em_{\omega}^2}{f_{\omega}}G_{\omega}+g_{\phi K^{*+}K^-}\frac{em_{\phi}^2}{f_{\phi}}RG_{\phi}) \ ,\\  g_{K^0K^{*0}\gamma}=&i(g_{\rho^0K^{*0}\bar{K}^{*0}}\frac{em_{\rho^0}^2}{f_{\rho^0}}G_{\rho^0}+g_{\omega K^{*0}\bar{K}^{*0}}\frac{em_{\omega}^2}{f_{\omega}}G_{\omega}+g_{\phi K^{*0}\bar{K}^{*0}}\frac{em_{\phi}^2}{f_{\phi}}RG_{\phi}) \ ,
   \\    g_{D^{*0}D^{0}\gamma}=&i(g_{\rho^0D^{*0}\bar{D}^{0}}\frac{em_{\rho^0}^2}{f_{\rho^0}}G_{\rho^0}+g_{\omega D^{*0}\bar{D}^{0}}\frac{em_{\omega}^2}{f_{\omega}}G_{\omega}+g_{\phi D^{*0}\bar{D}^{0}}\frac{em_{\phi}^2}{f_{\phi}}RG_{\phi}) \ ,
 \\    g_{D^{*+}D^{+}\gamma}=&i(g_{\rho^0D^{*+}D^{-}}\frac{em_{\rho^0}^2}{f_{\rho^0}}G_{\rho^0}+g_{\omega D^{*+}D^{-}}\frac{em_{\omega}^2}{f_{\omega}}G_{\omega}+g_{\phi D^{*+}D^{-}}\frac{em_{\phi}^2}{f_{\phi}}RG_{\phi}) \ ,
\\  g_{K^{*+}K^{*+}\gamma}=&i(g_{\rho^0K^{*+}K^{*-}}\frac{em_{\rho^0}^2}{f_{\rho^0}}G_{\rho^0}+g_{\omega K^{*+}K^{*-}}\frac{em_{\omega}^2}{f_{\omega}}G_{\omega}+g_{\phi K^{*+}K^{*-}}\frac{em_{\phi}^2}{f_{\phi}}RG_{\phi}) \ ,
\\
  g_{K^{*0}{K}^{*0}\gamma}=&i(g_{\rho^0K^{*0}\bar{K}^{*0}}\frac{em_{\rho^0}^2}{f_{\rho^0}}G_{\rho^0}+g_{\omega K^{*0}\bar{K}^{*0}}\frac{em_{\omega}^2}{f_{\omega}}G_{\omega}+g_{\phi K^{*0}\bar{K}^{*0}}\frac{em_{\phi}^2}{f_{\phi}}RG_{\phi}) \ ,
\\  g_{D^{*0}D^{*0}\gamma}=&i(g_{\rho^0D^{*0}D^{*0}}\frac{em_{\rho^0}^2}{f_{\rho^0}}G_{\rho^0}+g_{\omega D^{*0}D^{*0}}\frac{em_{\omega}^2}{f_{\omega}}G_{\omega}+g_{\phi D^{*0}D^{*0}}\frac{em_{\phi}^2}{f_{\phi}}RG_{\phi}) \ ,
\\
  f_{D^{*0}D^{*0}\gamma}=&4i (f_{\rho^0D^{*0}D^{*0}}\frac{em_{\rho^0}^2}{f_{\rho^0}}G_{\rho^0}+f_{\omega D^{*0}D^{*0}}\frac{em_{\omega}^2}{f_{\omega}}G_{\omega}+f_{\phi D^{*0}D^{*0}}\frac{em_{\phi}^2}{f_{\phi}}RG_{\phi}) \ ,
\\  g_{D^{*+}D^{*+}\gamma}=&i(g_{\rho^0D^{*+}D^{*-}}\frac{em_{\rho^0}^2}{f_{\rho^0}}G_{\rho^0}+g_{\omega D^{*+}D^{*-}}\frac{em_{\omega}^2}{f_{\omega}}G_{\omega}+g_{\phi D^{*+}D^{*-}}\frac{em_{\phi}^2}{f_{\phi}}RG_{\phi}) \ ,
\\  f_{D^{*+}D^{*+}\gamma}=&4i(f_{\rho^0D^{*+}D^{*-}}\frac{em_{\rho^0}^2}{f_{\rho^0}}G_{\rho^0}+f_{\omega D^{*+}D^{*-}}\frac{em_{\omega}^2}{f_{\omega}}G_{\omega}+f_{\phi D^{*+}D^{*-}}\frac{em_{\phi}^2}{f_{\phi}}RG_{\phi}) \ ,
\\  g_{K^{+}K^{+}\gamma}=&i(g_{\rho^0K^{+}K^{-}}\frac{em_{\rho^0}^2}{f_{\rho^0}}G_{\rho^0}+g_{\omega K^{+}K^{-}}\frac{em_{\omega}^2}{f_{\omega}}G_{\omega}+g_{\phi K^{+}K^{-}}\frac{em_{\phi}^2}{f_{\phi}}RG_{\phi}) \ ,\\  g_{K^{0}K^{0}\gamma}=&i(g_{\rho^0K^{0}\bar{K}^{0}}\frac{em_{\rho^0}^2}{f_{\rho^0}}G_{\rho^0}+g_{\omega K^{0}\bar{K}^{0}}\frac{em_{\omega}^2}{f_{\omega}}G_{\omega}+g_{\phi K^{0}\bar{K}^{0}}\frac{em_{\phi}^2}{f_{\phi}}RG_{\phi}) \ ,
\\  g_{D^{+}D^{+}\gamma}=&i(g_{\rho^0D^{+}D^{-}}\frac{em_{\rho^0}^2}{f_{\rho^0}}G_{\rho^0}+g_{\omega D^{+}D^{-}}\frac{em_{\omega}^2}{f_{\omega}}G_{\omega}+g_{\phi D^{+}D^{-}}\frac{em_{\phi}^2}{f_{\phi}}RG_{\phi})  \ ,\\  g_{D^{0}D^{0}\gamma}=&i(g_{\rho^0D^{0}\bar{D}^{0}}\frac{em_{\rho^0}^2}{f_{\rho^0}}G_{\rho^0}+g_{\omega D^{0}\bar{D}^{0}}\frac{em_{\omega}^2}{f_{\omega}}G_{\omega}+g_{\phi D^{0}\bar{D}^{0}}\frac{em_{\phi}^2}{f_{\phi}}RG_{\phi}) \ ,
\end{align}
where $R=0.8$ is the $SU(3)$ flavor symmetry breaking parameter for processes involving the strange $s$ quark.

\begin{figure}[!t]
  \begin{minipage}[t]{0.4\textwidth}
    \begin{center}
      \includegraphics[width=1\textwidth]{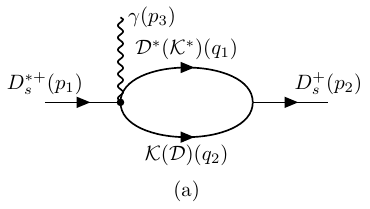}
          \end{center}
          \end{minipage}
           \begin{minipage}[t]{0.4\textwidth}
          \begin{center}
      \includegraphics[width=1\textwidth]{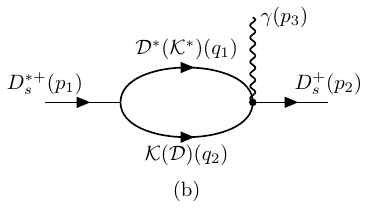}
          \end{center}
          \end{minipage}\\
           \begin{minipage}[t]{0.4\textwidth}
          \begin{center}
      \includegraphics[width=1\textwidth]{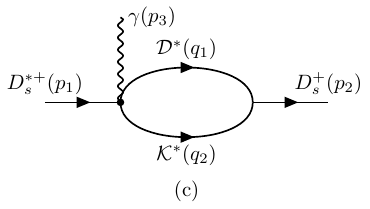}
          \end{center}
          \end{minipage}
           \begin{minipage}[t]{0.4\textwidth}
          \begin{center}
      \includegraphics[width=1\textwidth]{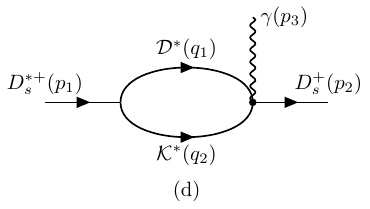}
          \end{center}
          \end{minipage}
          \caption{Contact diagrams of the decay $D^{*+}_{s}\to D^{+}_{s}\gamma$ .}\label{fig:contact}
          \end{figure}   
In addition to the contributions from Fig.~\ref{fig:vmdloop}, we also need to consider the contributions from the contact diagrams shown in Fig.~\ref{fig:contact}, which are induced by the requirement of Lorentz gauge invariance for the EM transitions. Similar to what found in Ref.~\cite{Li:2007xr}, we confirm that  Figs.~\ref{fig:contact}(b)-(d), do not contribute to the amplitude, as the integrals involve odd powers of momentum after the Feynman parameterization. Consequently, only Fig.~\ref{fig:contact}(a) has non-vanishing contributions to the amplitude.

The corresponding amplitudes arising from Fig.~\ref{fig:contact}(a) have the following expressions:
\begin{align}
  i \mathcal{M}_{a}(\mathcal{D}^{*},\mathcal{K},\gamma)
  =&-\int \frac{d^{4}q_{3}}{(2\pi)^{4}}\frac{g_{\mathcal{D}^{*}\mathcal{D}_{s}\mathcal{K}}g_{\mathcal{D}^{*}_{s}\mathcal{D}^{*}\mathcal{K}}\varepsilon_{\mu\nu\alpha \beta}\varepsilon^{s\mu}(p_1^{\nu} \varepsilon_{3}^{\alpha}+\varepsilon_{3}^{\nu}q_1^{\alpha} )p_2^{\rho}}{(q_1^{2}-m_1^{2})(q_2^{2}-m_2^{2})}
 \qty(g^{\beta \rho}-\frac{q_{1}^{\beta}q_{1}^{\rho}}{m_1^{2}})\mathcal{F}(q_{i}^{2}),
\\
  i \mathcal{M}_{a}(\mathcal{K}^{*},\mathcal{D},\gamma)
  =&-\int \frac{d^{4}q_{3}}{(2\pi)^{4}}\frac{4g_{\mathcal{D}\mathcal{D}_{s}\mathcal{K}^{*}}f_{\mathcal{D}^{*}_{s}\mathcal{D}\mathcal{K}^{*}}\varepsilon_{\mu\nu\alpha \beta}\varepsilon^{s\mu}(p_1^{\nu} \varepsilon_{3}^{\alpha}+\varepsilon_{3}^{\nu}q_1^{\alpha} )(p_2+q_2)^{\rho}}{(q_1^{2}-m_1^{2})(q_2^{2}-m_2^{2})}
 \qty(g^{\beta \rho}-\frac{q_{1}^{\beta}q_{1}^{\rho}}{m_1^{2}})\mathcal{F}(q_{i}^{2}) ,
\end{align}
where the same form factor as Eq.~(\ref{formfactor}) is adopted.

From the above amplitudes, we can obtain the total radiative decay loop amplitude by summing up the contributions from the different loop diagrams. The decay $D^{*+}_{s}\to D^{+}_{s}\gamma$ is a $VVP$ type decay process, and we can always parametrize the total loop amplitude as
 \begin{equation}
   i \mathcal{M}^{\gamma}_{\text{loop}}= i g_{\text{loop}}^{\gamma}\varepsilon_{\mu\nu\alpha \beta}p_{1}^{\mu}p_3^{\nu}\varepsilon^{\alpha}_{1}\varepsilon^{\beta}_{3}
  \end{equation} 
Taking into account the tree amplitude, the total decay amplitude can be expressed as
  \begin{equation}
   i \mathcal{M}_{D^{*+}_{s}\to D^{+}_{s}\gamma}=i(g_{\text{tree}}^{\gamma}+g_{\text{loop}}^{\gamma})\varepsilon_{\mu\nu\alpha \beta}p_{1}^{\mu}p_3^{\nu}\varepsilon^{\alpha}_{1}\varepsilon^{\beta}_{3}\equiv ig_{\text{total}}^{\gamma}\varepsilon_{D^{*+}_{s}}\cdot (p_{D^{+}_{s}}-p_{\pi^0}) \ ,
   \end{equation} 
where $g_{\text{total}}^{\gamma}$ is the effective coupling for $D^{*+}_{s}\to D^{+}_{s}\gamma$. Then, the corresponding partial decay width is
\begin{equation}
  \Gamma_{D^{*+}_{s}\to D_{s}^{+}\gamma}=  \frac{\abs{\mathbf{p}}^{3}(g^{\gamma}_{\text{total}})^{2}}{12\pi}.
\end{equation}

\section{Numerical Results and Discussions}

Proceeding to the numerical calculations of the tree-level and loop contributions for $D^{*+}_{s}\to D_{s}^+\pi^0$ and $D^{*+}_{s}\to D_{s}^{+}\gamma$, those common coupling constants are listed in Tables~\ref{tabcoupling1}-\ref{tabcoupling4}, with the relative phases determined by SU(4) flavor symmetry. This advantage allows us to combine these two decay processes together, and investigate the experimental constraint on the range of the only individual parameter $g_{J/\psi D\bar{D}}$ in $D^{*+}_{s}\to D_{s}^{+}\gamma$. Note that the same form factor is introduced in the loop integrals. We can also investigate the dependence of the branching ratio fraction between these two decay channels on the cut-off parameter. The masses of the relevant particles are taken from the PDG \cite{ParticleDataGroup:2024cfk}.

\begin{table}[!ht]
     \centering\caption{The values of the $VPP$ coupling constants.}
     \begin{ruledtabular}   
     \begin{tabular}{cccccccccc}
          Coupling Constant & $g_{D_{s}^{*+}D^{0}K}$& $g_{D_{s}^{+}D^{0}K^*}$& $g_{D^{*0}D^{0}\pi}$&$g_{D^{*-}D^{+}\pi}$ &$g_{D_{s}^{*+}D^{+}K}$&  $g_{D^{*0}D_{s}^{+}K}$&$g_{D_{s}^{+}D^{+}K^*}$&$g_{D^{*+}D_{s}^{+}K}$&  $g_{VPP}$\\\hline
      Numerical Value & $18.40$&$3.84 $&$17.29$&$-17.33$&$18.42$&$17.77$& $3.84$&$17.78$&$4.18$
     \end{tabular}
   \end{ruledtabular}\label{tabcoupling1}
  \end{table}
   \begin{table}[!ht]
     \centering\caption{The values of the $VVP$ coupling constants.}
     \begin{ruledtabular}
   \begin{tabular}{ccccccc}
         Coupling Constant&$g_{D_{s}^{*+}D^{*0}K}$ &$f_{D^{*0}D^{+}_{s}K^{*}}$&$f_{D^{*}_{s}D^{0}K^{*}}$&$g_{D^{*0}\bar{D}^{*0}\pi^0}$&$g_{D^{*+}D^{*-}\pi^0}$&$g_{VVP}$\\\hline
         Numerical Value$(\mathrm{GeV}^{-1})$& $7.81$&$2.38$&$2.47$&$8.94$&$-8.94$&$7.93$
           \end{tabular}
         \end{ruledtabular}\label{tabcoupling2}
 \end{table}
 \begin{table}[!ht]
   \centering\caption{The values of the $VVV$ coupling constants.}
   \begin{ruledtabular}
  \begin{tabular}{cccc}
     Coupling Constant&$g_{VVV}$&$g_{D^{*+}_{s}D^{*0}K^{*}}$&$f_{D^{*}_s D^{*0} K^{*}}$\\\hline
   Numerical Value& 4.47 & 3.83&4.79
       \end{tabular}
     \end{ruledtabular}\label{tabcoupling3}
 \end{table}
 \begin{table}[!ht]
  \centering\caption{The values of the electromagnetic coupling constants in VMD model.}
  \begin{ruledtabular}
 \begin{tabular}{ccccc}
    Coupling Constant&$g_{K^{*+}K^{+}\gamma}$&$g_{K^{*0}K^{0}\gamma}$&$g_{D^{*0}D^{0}\gamma}$&$g_{D^{*+}D^{+}\gamma}$\\\hline
  Numerical Value($\mathrm{GeV}^{-1})$& $-0.288+0.063i$&$0.369+0.062i$&$-0.383-0.082i$&$0.492+0.082i$\\\hline
     Coupling Constant&$g_{K^{*+}K^{*+}\gamma}$&$g_{K^{*0}K^{*0}\gamma}$&$g_{D^{*0}D^{*0}\gamma}(f_{D^{*0}D^{*0}\gamma})$&$g_{D^{*+}D^{*+}\gamma}(f_{D^{*+}D^{*+}\gamma})$\\\hline
  Numerical Value& $-0.162-0.035i$&$0.208+0.034i$&$-0.139-0.031i(-0.695-0.152i)$&$0.178+0.030i(0.892+0.150i)$\\\hline
     Coupling Constant&$g_{K^{+}K^{+}\gamma}$&$g_{K^{0}K^{0}\gamma}$&$g_{D^{0}D^{0}\gamma}$&$g_{D^{+}D^{+}\gamma}$\\\hline
  Numerical Value& $-0.288+0.063i$&$0.369+0.062i$&$-0.383-0.082i$&$0.492+0.082i$
      \end{tabular}
    \end{ruledtabular}\label{tabcoupling4}
\end{table}

In Table~\ref{tabresult1} we list the calculation results of $D^{*+}_{s}\to D_{s}^{+}\pi_0$ for the tree, loop transitions and the total, respectively, with five typical cut-off parameter values $\alpha=1.0$, $1.35$, $1.5$, $1.65$ and $2.0$. We can see that although there exist some sensitivities of the loop contribution to the cutoff parameter $\alpha$, the loop contributions turn out to be rather stable. Combining the tree and loop amplitudes together, the partial decay width reads $\Gamma_{\text{total}} = 9.92^{+0.76}_{-0.66}\,\mathrm{eV}$ in the range of $\alpha=1.5\pm 0.15$, which serves as an reasonable estimate of the model uncertainties in our theoretical calculations. Similar to what found in Ref.~\cite{Yang:2019cat} that the small higher-order corrections can introduce rather significant interferences in the final results, we also find that the loop contributions cannot be neglected.

 \begin{table}[!ht]
     \centering\caption{Contributions of the tree diagram, loop diagrams, and the combination of tree and loop diagrams to the partial decay width of $D^{*+}_{s}\to D_{s}^{+}\pi^{0}$ with $\alpha=1.0$, $1.35$, $1.5$, $1.65$ and $2.0$ in unit of $\mathrm{eV}$.}
       \begin{ruledtabular}
   \begin{tabular}{cccccc}
         $\alpha$&$1.0$ &$1.35$&$1.5$&$1.65$&$2.0$\\\hline
      $\Gamma_{\text{tree}}$& $6.93$&$6.93$&$6.93$&$6.93$&$6.93$\\
      $\Gamma_{\text{loop}}$&$0.04$&$0.17$&$0.26$&$0.41$&$0.90$\\
      $\Gamma_{\text{total}}$&$8.07$&$9.26$&$9.92$&$10.67$&$12.81$
           \end{tabular}
             \end{ruledtabular}\label{tabresult1}
 \end{table}
 
In Table~\ref{tabcompare} we compare our result with other theoretical calculations, i.e. the covariant model (CM)~\cite{Cheung:2015rya} and the chiral perturbation theory ($\chi$PT)~\cite{Yang:2019cat}. It shows that Our result is consistent with that obtained from $\chi$PT~\cite{Yang:2019cat}. Note that the results from the CM~\cite{Cheung:2015rya} are significantly larger than both our result and the $\chi$PT calculation~\cite{Yang:2019cat}.
 \begin{table}[!ht]
   \centering
   \caption{Comparisons of the partial decay width of $D^{*+}_{s}\to D^{+}_{s}\pi^0$. The decay widths are in the uints of $\mathrm{eV}$. The uncertainties of our result are given by $\alpha=1.5\pm 0.15$.}
   \begin{ruledtabular}
 \begin{tabular}{cccc}
       &CM~\cite{Cheung:2015rya}& $\chi$PT~\cite{Yang:2019cat}&Our work\\\hline
    $\Gamma(D^{*+}_{s}\to D^{+}_{s}\pi^0)$& $277^{+28}_{-26}$&$8.1^{+3.0}_{-2.6}$&$9.92^{+0.76}_{-0.66}$\\ 
 \end{tabular}
 \end{ruledtabular}\label{tabcompare}
 \end{table}
 
\begin{figure}
     \centering
       \includegraphics[width=0.65\textwidth]{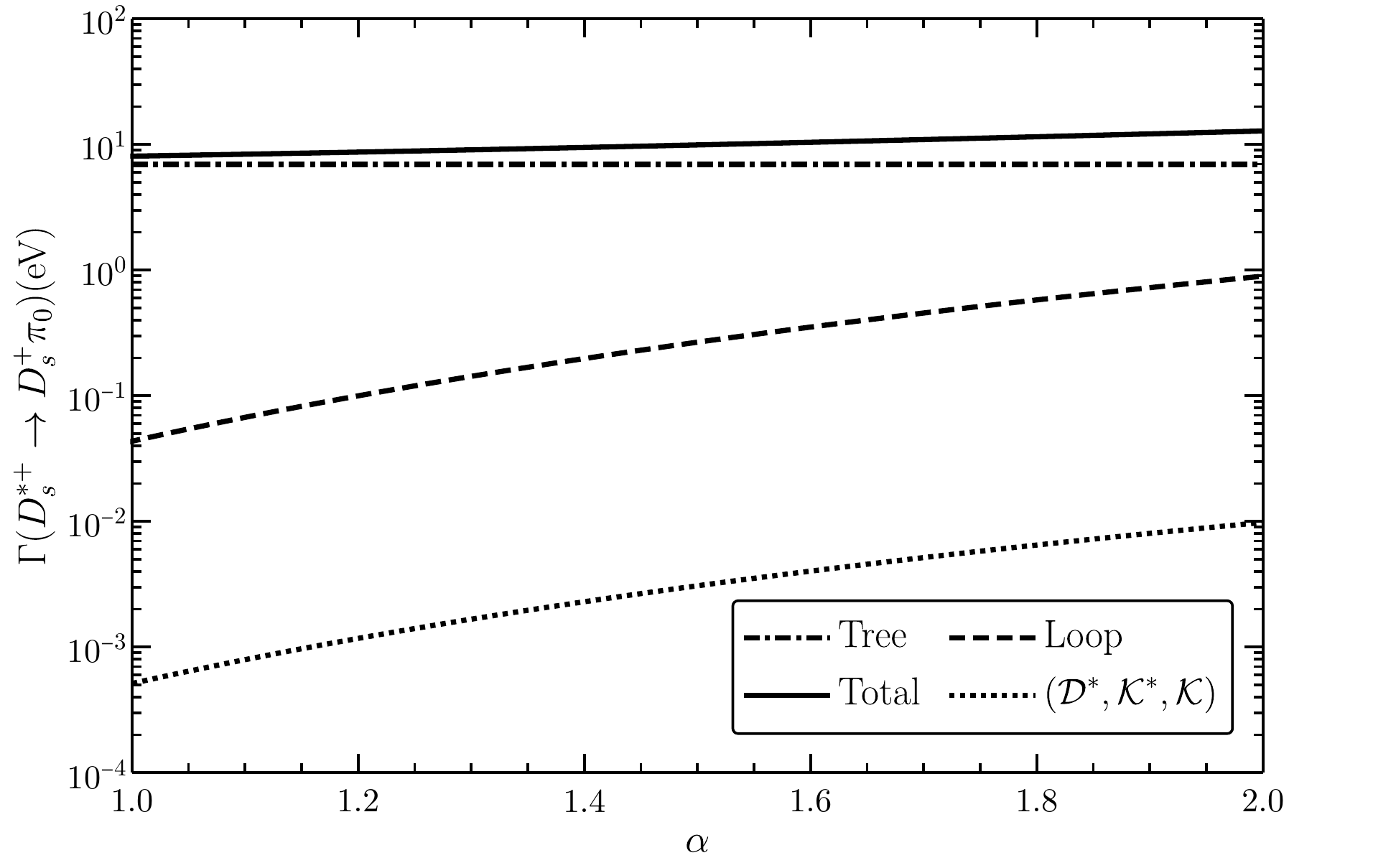}
       \caption{The partial decay width of $D^{*+}_{s}\to D_{s}^{+} \pi^0$ is shown as a function of the cut-off parameter $\alpha$. The solid curve illustrates the total contribution, the dot-dashed curve indicates the tree-level contribution, and the dashed curve shows the loop contributions. The dotted curve depicts the contribution from the $(\mathcal{D}^{*},\mathcal{K}^{*},\mathcal{K})$ loop as a demonstration of the contribution from a single process.}\label{figdecaywidth}
       \end{figure}

To see more clearly the role played by the intermediate loop transitions, we plot the dependence of the decay width of $D^{*+}_{s}\to D_{s}^{+} \pi^0$ on the cut-off parameter $\alpha$, which includes contributions from tree diagrams, loop diagrams, and the sum of tree and loop diagrams, as shown in Fig.~\ref{figdecaywidth}. It should be noted that the contributions from the intermediate meson loops are cumulative, with different channels contributing comparably. The channel involving the exchange of $\mathcal{K}$, i.e. the $(\mathcal{D}^{*},\mathcal{K}^{*},\mathcal{K})$ channel, has a relatively larger contribution, approximately twice of the other channels. To better illustrate the cumulative effect, we plot the exclusive contribution of this channel in Fig.~\ref{figdecaywidth}. As shown there, this relatively dominant loop transition only contributes a small portion of the total width.

As discussed earlier, the transitions $D^{*+}_{s}\to D_{s}^{+} \pi^0$ and $D^{*+}_{s}\to D_{s}^{+}\gamma$ share the same strong couplings except for  $g_{J/\psi D\bar{D}}$ which appears in the tree amplitude of $D^{*+}_{s}\to D_{s}^{+}\gamma$. In the literature one finds that this coupling takes values in a range of $g_{J/\psi D\bar{D}}=7\sim 7.5$~\cite{Lin:1999ad,Matheus:2002nq,Oh:2000qr,Oh:2007ej,Zhang:2009kr}. As we will show later that the loop correction in $D^{*+}_{s}\to D_{s}^{+}\gamma$ is negligibly small, it thus allows us to conclude that the main uncertainty source in this combined analysis should come from the tree-level contribution, i.e. the uncertainty of $g_{J/\psi D\bar{D}}$. 

Our strategy below is to first constrain the coupling  $g_{J/\psi D\bar{D}}$ using the experimental data for the branching ratio fraction of $D^{*+}_{s}\to D_{s}^{+}\pi^{0}$ to 
$D^{*+}_{s}\to D_{s}^{+}\gamma$~\cite{ParticleDataGroup:2024cfk,BESIII:2022kbd}. We then constrain $g_{J/\psi D\bar{D}}$ by assuming that this coupling is the only adjustable parameter for reproducing the branching ratio fraction with the cut-off parameter $\alpha=1.5$, while all the other common couplings are fixed by $D^{*+}_{s}\to D_{s}^{+}\pi^{0}$. With the branching ratio fraction data from the PDG~\cite{ParticleDataGroup:2024cfk} we extract:
\begin{equation}
  g_{J\psi D \bar{D}}=7.23 \pm 0.06.
 \end{equation} 
This value is within the range of $g_{J/\psi D\bar{D}}=7\sim 7.5$ adopted in the literature~\cite{Lin:1999ad,Matheus:2002nq,Oh:2000qr,Oh:2007ej,Zhang:2009kr}, but is better constrained.

       \begin{table}[!ht]
          \centering\caption{Experimentally measured branching ratios and relative branching ratios of $D^{*+}_{s}\to D_{s}^{+}\pi^{0}$ and $D^{*+}_{s}\to D_{s}^{+}\gamma$.}
               \begin{ruledtabular}
                 \begin{tabular}{cccc}
               &$\mathrm{BR}(D^{*+}_{s}\to D^{+}_{s}\gamma)$&$\mathrm{BR}(D^{*+}_{s}\to D^{+}_{s}\pi^{0})$&$\mathrm{BR}(D^{*+}_{s}\to D^{+}_{s}\pi^{0})/\mathrm{BR}(D^{*+}_{s}\to D^{+}_{s}\gamma)$\\\hline
           PDG \cite{ParticleDataGroup:2024cfk}&$(93.6\pm 0.4)\%$ &$(5.77 \pm 0.35)\%$&$(6.2\pm 0.4)\%$
         \\
         BESIII \cite{BESIII:2022kbd}&$(93.54\pm 0.38\pm 0.22)\%$&$(5.76 \pm 0.38\pm 0.16)\%$&$(6.16\pm 0.43\pm 0.18)\%$
               \\
                 \end{tabular}
               \end{ruledtabular}\label{tabresult2}
       \end{table}
       \begin{table}[!ht]
        \centering\caption{The partial decay width of $D^{*+}_{s}\to D_{s}^{+}\gamma$ and the total width of $D^{*+}_{s}$ obtained using the experimentally measured relative branching ratio and the calculated width of $D^{*+}_{s}\to D_{s}^{+}\pi^{0}$ with $\alpha=1.5$.}
               \begin{ruledtabular}
                 \begin{tabular}{ccc}
              &$\Gamma(D^{*+}_{s}\to D_{s}^{+}\gamma)$&$\Gamma_{\text{total}}(D^{*+}_{s})$\\\hline
             PDG \cite{ParticleDataGroup:2024cfk}&$160^{+10}_{-10}\mathrm{eV}$& $172^{+10}_{-10}\,\mathrm{eV}$
       \\
       BESIII \cite{BESIII:2022kbd}&$161^{+16}_{-16}\mathrm{eV}$&$172^{+16}_{-16}\,\mathrm{eV}$
             \\
               \end{tabular}
             \end{ruledtabular}\label{tabresult3}
     \end{table}

       \begin{table}[!ht]
        \centering\caption{Contributions of the tree diagram, loop diagrams, and the combination of tree and loop diagrams to the partial decay width of $D^{*+}_{s}\to D_{s}^{+}\gamma$ with $\alpha=1.0$, $1.35$, $1.5$, $1.65$ and $2.0$ in unit of $\mathrm{eV}$ and $g_{J/ \psi D\bar{D}}=7.23$.}
          \begin{ruledtabular}
      \begin{tabular}{cccccc}
            $\alpha$&$1.0$ &$1.35$&$1.5$&$1.65$&$2.0$\\\hline
         $\Gamma_{\text{tree}}$& $153.89$&$153.89$&$153.89$&$153.89$&$153.89$\\
         $\Gamma_{\text{loop}}$&$0.54$&$2.22$&$3.59$&$5.55$&$13.14$\\
         $\Gamma_{\text{total}}$&$155.30$&$157.84$&$159.69$&$162.16$&$171.15$
              \end{tabular}
                \end{ruledtabular}\label{tabresult4}
    \end{table}
    
    \begin{table}[!ht]
      \centering\caption{The partial decay widths of $D^{*+}_{s} \to D^{+}_{s} \pi^{0}$ and $D^{*+}_s\to D^{+}_{s} \gamma$, as well as the total width of $D^{*+}_{s}$ as the sum of both, for $\alpha=1.5\pm 0.15$ with $g_{J\psi D \bar{D}}$ fixed at $7.23$, and for $g_{J\psi D \bar{D}}=7.23\pm 0.06$.}
        \begin{ruledtabular}
    \begin{tabular}{lccc}
     &$\Gamma(D^{*+}_{s}\to D_{s}^{+}\pi^{0})$&$\Gamma(D^{*+}_{s}\to D_{s}^{+}\gamma)$&$\Gamma_{\text{total}}(D^{*+}_{s})$\\\hline
     $\alpha=1.5\pm 0.15$, $g_{J_{\psi D \bar{D}}}=7.23$&$9.92^{+0.76}_{-0.66}$&$159.7^{+2.5}_{-1.8}$&$169.6^{+2.6}_{-2.0}$\\
     $\alpha=1.5\pm 0.15$, $g_{J_{\psi D \bar{D}}}=7.23 \pm 0.06$&$9.92^{+0.76}_{-0.66}$&$160^{+13}_{-12}$&$170^{+ 13}_{-12}$\\
            \end{tabular}
              \end{ruledtabular}\label{tabresult5}
  \end{table}

\begin{figure}
      %[!ht] 
           \centering
             \includegraphics[width=0.65\textwidth]{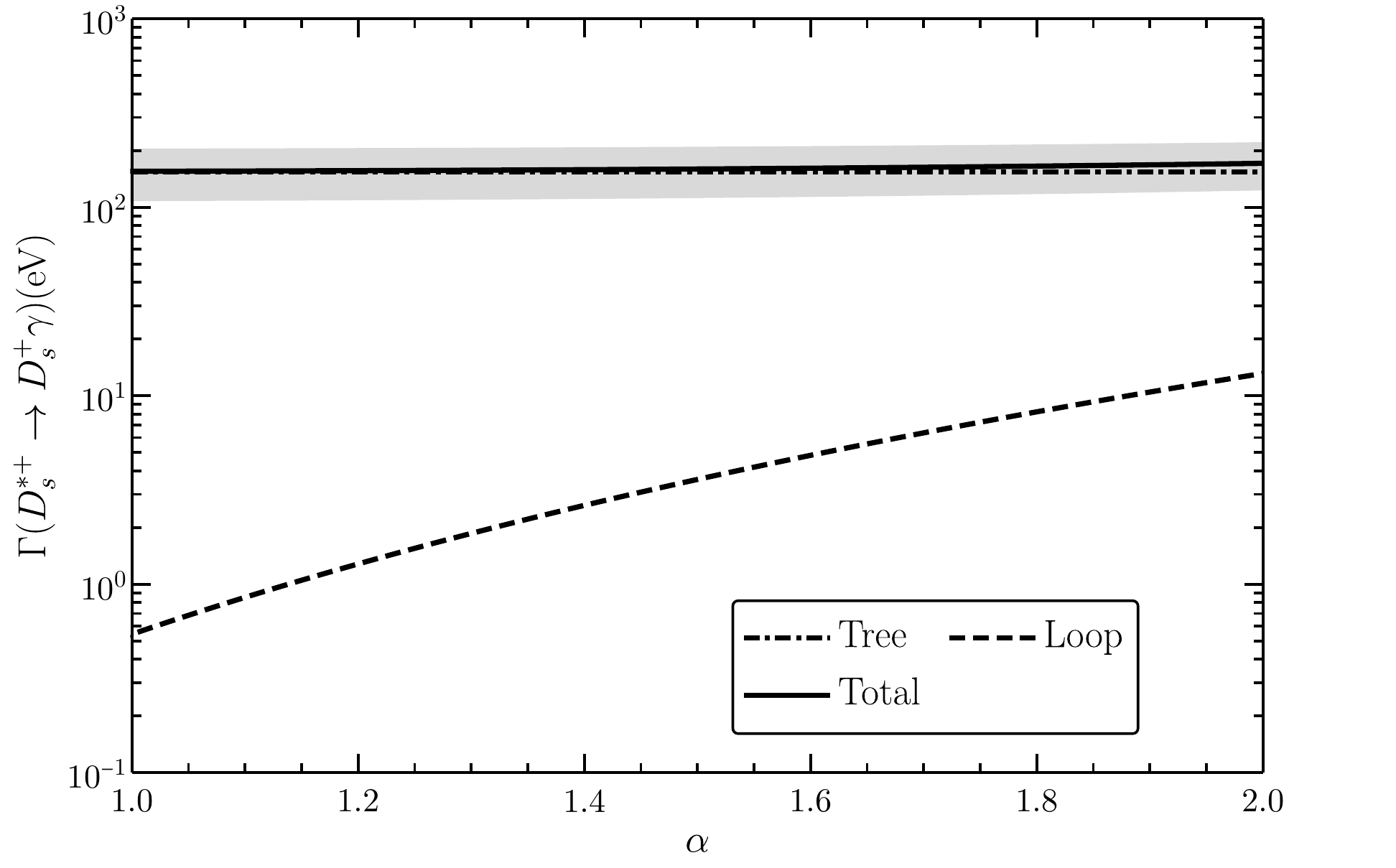}
            \caption{The partial decay width of $D^{*+}_{s}\to D_{s}^{+}\gamma$ is shown as a function of the cut-off parameter $\alpha$ with $g_{J/\psi D\bar{D}}=7.23$. The solid curve illustrates the total contribution, the dot-dashed curve indicates the tree-level contribution, and the dashed curve shows the loop contributions. The light gray band represents the variation range of the results for $g_{J/\psi D\bar{D}}=7\sim 7.5$.}\label{figgammawidth}
             \end{figure}

It is necessary to examine the magnitude of the loop corrections for $D^{*}_{s}\to D_s\gamma$ and its sensitivity to the cut-off parameter $\alpha$. 
In Table \ref{tabresult4}, we list the calculated partial decay widths of $D^{*}_{s}\to D_s\gamma$ for the tree-level, loop contributions, and the total, with five typical cut-off parameter values $\alpha=1.0$, $1.35$, $1.5$, $1.65$, and $2.0$, and $g_{J\psi D \bar{D}}=7.23$. In Fig.~\ref{figgammawidth}, we plot the cut-off dependence of the partial decay width of $D^{*+}_{s}\to D_{s}^{+}\gamma$, as well as the variation range of the results for $g_{J\psi D \bar{D}}=7\sim 7.5$. In contrast with the $D^{*}_{s}\to D_{s}\pi^{0}$ decay channel, the loop corrections to the tree-level amplitude are negligibly small.
The uncertainty in the calculation of the  partial width for $D^{*}_{s}\to D_{s}\gamma$ is predominantly determined by the tree-level contributions. The error in the tree-level contributions mainly arises from the uncertainty of $g_{J\psi D \bar{D}}$. Notice that within the range $g_{J/\psi D\bar{D}}=7\sim 7.5$, the partial width of $D^{*}_{s}\to D_{s}\gamma$ varies approximately between $100 \sim 200 \,\mathrm{eV}$, which encompasses the result estimated using the decay width of $D^{*}_{s}\to D_{s}\pi^{0}$. This can be regarded as a self-consistency check of the formalism.

To examine the dependence of the partial decay width and total width of $D^{*+}_{s}$ on the cut-off parameter $\alpha$ and the coupling constant $g_{J\psi D \bar{D}}$, we present in Table~\ref{tabresult5} the partial decay widths of $D^{*+}_{s} \to D^{+}_{s} \pi^{0}$ and $D^{*+}_s\to D^{+}_{s} \gamma$, as well as the total width of $D^{*+}_{s}$ as the sum of both, for $\alpha=1.5\pm 0.15$ with $g_{J\psi D \bar{D}}$ fixed at $7.23$, and for $g_{J\psi D \bar{D}}=7.23\pm 0.06$, respectively. Note that $\Gamma(D_s^{*+}\to D_s^+\pi^0)$ does not depend on $g_{J\psi D \bar{D}}$. The variation of $g_{J\psi D \bar{D}}$ will only bring errors to $\Gamma(D^{*+}_{s}\to D_{s}^{+}\gamma)$, and this absorbs the primary uncertainties for the estimate of the total width.

It should be stressed that our formalism provides a relatively strigent constraint on the coupling $g_{J\psi D \bar{D}}$ which results in an improved etimate of the total width. Recalling that the range of $g_{J/\psi DD}=7\sim 7.5$ is adopted in the literature~\cite{Lin:1999ad,Matheus:2002nq,Oh:2000qr,Oh:2007ej,Zhang:2009kr}, with such a large uncertainty for $g_{J\psi D \bar{D}}$, the uncertainties with the total width can amount to more than $20\%$. In this sense we may regard that the decay of $D_s^*$ can provide a reliable constraint on the coupling $g_{J\psi D \bar{D}}$. 
 
\section{Summary}
In this work we present a combined study of the isospin-violating decay $D_s^* \to D_s \pi^0$ and radiative decay $D_s^*\to D_s\gamma$. The isospin-violating decay $D_s^* \to D_s \pi^0$ is investigated with both the tree-level $\eta-\pi^0$ mixing and higher-order meson loop corrections from the intermediate $D^{({*})}$ and $K^{({*})}$ rescatterings. The radiative decay $\Gamma(D^{*+}_{s}\to D_{s}^{+}\gamma)$ is investigated in the same framework via the VMD model. Our findings indicate that the tree-level contributions are dominant in both processes, while loop effects play a crucial role in the understanding of the $D^{*}_{s} \to D_{s}\pi^{0}$ decay. In particular, the $D^{*}K^{*}$ rescatterings by exchange a $K$ account for most of the loop corrections. In contrast, the loop corrections in $\Gamma(D^{*+}_{s}\to D_{s}^{+}\gamma)$ are negligibly small. With the experimental data for the branching ratio fraction, we obtain a better constraint on the coupling $g_{J/\psi D\bar{D}}$, which is crucial for a reliable estimate of the radiative decay width. These mechanisms can also help us understand other isospin-violating decay processes. Since the decay $D^{*}_{s}\to D_{s}\pi_0$ is near the threshold of $D_{s}^{*}$, understanding this decay channel will aid in comprehending other near-threshold dynamics. Future precise measurement of the total width of $D^{*}_{s}$ at BESIII is strongly recommended.
       
\section*{Acknowledgement}
Useful discussion with Prof. Shi-Lin Zhu is acknowledged. This work is supported, in part, by the National Natural Science Foundation of China (Grant No. 12235018), DFG and NSFC funds to the Sino-German CRC 110 ``Symmetries and the Emergence of Structure in QCD" (NSFC Grant No. 12070131001, DFG Project-ID 196253076), National Key Basic Research Program of China under Contract No. 2020YFA0406300, and Strategic Priority Research Program of Chinese Academy of Sciences (Grant No. XDB34030302).
\bibliographystyle{unsrt}
\bibliography{Dsstar_Decay.bib}
\end{document}